\documentclass[%
 aip,
amsmath,
 amssymb,
 reprint,%
]{revtex4-1}
\usepackage{mathrsfs}
\usepackage{amsmath}
\usepackage{mathtools}

\newcommand{\angstrom}{\textup{\AA}}

\usepackage{framed,multirow}
\usepackage{graphicx}
\usepackage{dcolumn}
\usepackage{bm}

\usepackage{siunitx}

\usepackage[utf8]{inputenc}
\usepackage[T1]{fontenc}
\usepackage{mathptmx}
\DeclareMathAlphabet{\mathcal}{OMS}{cmsy}{m}{n}
\begin{document}

\preprint{AIP/123-QED}

\title[An Efficient Algorithm for Interfacial Statistical Associating Fluid (iSAFT) in Cylindrical Geometry]{An Efficient Algorithm for Interfacial Statistical Associating Fluid (iSAFT) in Cylindrical Geometry}

\author{Shun Xi}
\author{Jinlu Liu}
\author{Arjun Valiya Parambathu}
\author{Yuchong Zhang}
\author{Walter Chapman}
\email{wgchap@rice.edu}
\affiliation{Department of Chemical and Biomolecular Engineering, Rice University, 6100 Main St, Houston, TX 77005, USA
}%

\date{\today}

\begin{abstract}
In this work we present an efficient numerical algorithm for the solution of interfacial statistical associating fluid theory (iSAFT) in cylindrical geometry to facilitate the study of inhomogeneous fluids having curvatures. The new solution algorithm is shown to have a better time scaling than the elliptic function method by Malijevsky~\cite{malijevsky2007fundamental}, and the transform method by Lado~\cite{lado1971numerical}. Convergence, performance, and stability of the numerical algorithm are discussed. We showcase two representative applications of the new method for modeling fluid adsorption and bottlebrush polymers. By comparing iSAFT with molecular simulation results, we found that iSAFT predicts layering transitions above the triple point for methane adsorption, and it captures power-law to parabolic transitions for polymers brush microstructure. We conclude that the continuous development of solution algorithm for iSAFT enables researchers to investigate curvature effects for fluids in efficient manners.

\end{abstract}

\maketitle
\section{Introduction}
In recent years classical density functional theory (DFT) has been successfully applied to study the microscopic structure of inhomogeneous fluids after the successful introduction of fundamental measure theory (FMT) by Rosenfeld~\cite{rosenfeld1989free}, and to model complex polymeric fluids with the extension of chain connectivity using Wertheim's theory~\cite{wu2006density, emborsky2011recent}. When a fluid experiences an external field, inhomogeneous phenomena such as fluid adsorption, capillary condensation, self-assembly, etc., are observed. While density functional theory accurately predicts the equilibrium fluid structure with a three-dimensional degree of freedom, it is prevalent to set the fluid distribution as one-dimensional in space by assuming density symmetry, for example, fluid adsorbed in a slit-pore, cylindrical pore, spherical cavity, and polymer tethered to a planar wall, cylindrical nano-rod and spherical nano-particle. Mathematical manipulations are necessary to reduce density functional theory from a three-dimensional formulation to a one-dimensional formulation. The planar geometry DFT formulation can be easily obtained since the vectorial notations in weighted densities remain unchanged in Cartesian coordinate. The polar geometry formulations such as cylindrical and spherical situations are more involved~\cite{roth2010fundamental}. DFT in spherical geometry can be formulated using elementary functions~\cite{wang2017modeling}. However, cylindrical DFT formulation requires successive calculations of elliptic functions of the first and second kinds as provided by Malijevsky~\cite{malijevsky2007fundamental}. The implementation of the cylindrical DFT formulation by direct integration is thus non-trivial.

The Fourier transform method becomes an alternative route to solve density functional theory. The convolutional weighted densities and functional derivatives in DFT can be efficiently calculated in Fourier space. An early attempt in applying the Fourier transform to solve density functional theory was studied by Sears and Frink~\cite{sears2003new}. However, different numerical challenge arises in polar spherical geometry and cylindrical geometry, which requires additional manipulation to re-represent weighted densities. The Fourier transform of DFT in spherical geometry can be completed by using the Fourier sine transform. The Fourier transform of DFT in cylindrical geometry, however, has to be completed by the zeroth order Hankel transform.

The Hankel transform naturally appears in many physical problems when the Fourier transform is applied to a cylindrically symmetric system. In optical theory, the Hankel transform provides a method to solve the paraxial wave equation which describes the dynamics of cylindrical electromagnetic waves such as a laser beam~\cite{grella1982fresnel}. In quantum mechanics, the Hankel transform is the key part in the pseudo-spectral method~\cite{gottlieb1977numerical}, which effectively calculates the radial part of the Laplacian operator of the Schr\"{o}dinger equation, and it provides a reduction in grid size compared with the finite difference method~\cite{bisseling1985fast}. In statistical mechanics of equilibrium fluids, the Hankel transform can be applied to solve convolution-type integral equations by algebraically solving the transformed integral equation in Fourier space~\cite{lado1971numerical}. A classical example is the Ornstein{--}Zernike equation that defines the direct correlation function. Lado~\cite{lado1971numerical} provides a finite domain numerical implementation of Hankel transform by imposing a vanishing condition at the boundary. This method requires a matrix kernel to complete the transform. Therefore it is not an efficient implementation compared to the standard Fast Fourier Transform (FFT) method. Siegman provides a quasi-fast Hankel transform (QFHT) by applying an exponential transformation~\cite{gardner1959method,siegman1977quasi}. Magni et al. modified Siegman's formulation and provided a fast Hankel transform with high accuracy~\cite{magni1992high}. This formulation only requires vector kernel to complete the transform in which an efficient FFT algorithm can be applied. In this work, we extend the use of fast Hankel transform as an efficient algorithm for the solution of interfacial statistical associating fluid theory.

The original density functional theory that attempts to model spherical Lennard--Jones (LJ) molecules is limited by its the approximated mean-field free energy functional, and by its lack of molecular chain connectivity. Interfacial statistical associating fluid theory (iSAFT) a particular classical density functional theory by Tripathi and Chapman\cite{tripathi2005microstructure}, and Jain et al\cite{jain2007modified} is an extension of statistical associating fluid theory (SAFT)\cite{chapman1989saft} for inhomogeneous fluid. It has the advantage of modeling complex molecules beyond spherical shapes by including short-ranged association for water and alcohol\cite{ballal2013hydrophobic}, homogeneous chain connectivity for hydrocarbons\cite{liu2017adsorption}, heterogeneous chain connectivity for block copolymers\cite{emborsky2011exploring}, and branching effect for dendrimers\cite{zhang2018density} and star polymers\cite{bymaster2010saft}.
	
The previous studies using iSAFT focus on systems in planar geometry. However, curvature effect is important to understand the equilibrium properties of inhomogeneous fluids at nano-scale and meso-scale. The vapor pressure of forming a spherical nano-droplet depends on both surface tension and the droplet radii\cite{fisher1981kelvin}. Roth et al\cite{roth1999depletion} used density functional theory showed that the depletion forces near curved surfaces strongly deviates from the flat wall limit. The curvature effect has to be considered for modeling forces exerted onto the colloidal particles enclosed in vesicles. For the hard-sphere fluids having curved interfaces, the surface tensions are shown proportional to the logarithm of the radius of interface curvature by using density functional theory\cite{bryk2003hard}. Hlushak used density functional theory of hard-sphere 2-Yukawa model to study the optimal pore sizes for gas storage in fluid adsorption. It is found that the optimal pore size of cylindrical pore is significantly more dependent of bulk pressure than slit pore\cite{hlushak2018heat}. For long-chain polymers, Binder et al\cite{binder2017polymer} reported that cylindrical polymer brushes have two distinctive structures at low and high cylinder radius compared to flat polymer brushes. Therefore it is necessary to continuously extend the applications of iSAFT to curved systems. A few interesting systems of inhomogeneous fluids under spherical geometry have been investigated using iSAFT\cite{bymaster2008microstructure, wang2017modeling, xi2019thermodynamics}. In this work we developed an efficient algorithm for iSAFT in cylindrical geometry as a tool for more specialized problems. We detail two applications of iSAFT in cylindrical geometry. The first case is adsorption of methanes in a cylindrical pore with Steel potential. The second case is bottlebrush polymers in implicit solvents. Given the versatile of iSAFT for accurately modeling spherical molecules and chain molecules, the development of efficient solution algorithm for iSAFT will lead to further studies of complex fluids under cylindrical geometry.

The paper is outlined as follows: Section II describes the theoretical framework of iSAFT free energy functional. Section III details the implementation of the algorithm. Section IV verifies the algorithm, and discusses the performance, convergence, and stability of the new algorithm. Section V showcase how it can be applied to modeling fluid adsorption in cylindrical nano-pores and to modeling bottlebrush polymers.

\section{iSAFT Free Energy Functional}
\label{theory}
Density functional theory states that the equilibrium density distribution of inhomogeneous fluids is determined by the minimization of grand potential:
\begin{equation*}
\Omega[\rho] = A[\rho] - \sum_{\alpha} \int d\mathbf{r}\rho_{\alpha}(\mathbf{r})\left[\mu_{\alpha} - V^{ext}_{\alpha}(\mathbf{r})\right].
\end{equation*}
The minimization of grand potential produces the Euler--Lagrange equation:
\begin{equation}
\frac{\delta\Omega[\rho]}{\delta\rho_\alpha(\textbf{r})} = \frac{\delta A[\rho]}{\delta\rho_\alpha(\textbf{r})} - (\mu_\alpha-V_\alpha^{ext}(\textbf{r})) = 0
\label{eq:euler-lagrange}
\end{equation}
This is solved iteratively to obtain the equilibrium density distribution. $\Omega[\rho]$ and $A[\rho]$ are the total grand potential and Helmholtz free energy functional of the system, respectively. $\rho_{\alpha}(\mathbf{r})$ is the spatial density distribution of component $\alpha$. $\mu_\alpha$ is the chemical potential of component $\alpha$. $V_\alpha^{ext}(\textbf{r})$ is the external potential for component $\alpha$.

The free energy functional of iSAFT is based on perturbation theory that the Helmholtz free energy can be decomposed into reference fluid and perturbation: 
\begin{equation*}
A[\rho] = A^{id}[\rho] + A^{hs}[\rho] + A^{disp}[\rho] + A^{ch}[\rho],
\end{equation*}
where $ A^{id}[\rho] $, $A^{hs}[\rho]$, $A^{disp}[\rho]$, $A^{ass}[\rho]$, and $A^{ch}[\rho]$ are free energy functional of ideal gas contribution, hard-sphere contribution, dispersion contribution, association and chain contribution. The Euler--Lagrange equation~(\ref{eq:euler-lagrange}) under the framework of perturbation theory is:
\begin{align}
\frac{\delta A^{id}[\rho]}{\delta\rho_\alpha(\textbf{r})}+\frac{\delta A^{hs}[\rho]}{\delta\rho_\alpha(\textbf{r})} + \frac{\delta A^{disp}[\rho]}{\delta\rho_\alpha(\textbf{r})}+ \frac{A^{ass}[\rho]}{\delta\rho_\alpha(\textbf{r})} + \frac{A^{ch}[\rho]}{\delta\rho_\alpha(\textbf{r})} = (\mu_\alpha-V_\alpha^{ext}(\textbf{r})).
\end{align}


The ideal gas functional is:
\begin{equation*}
A^{id}[\rho] = k_BT \sum_{\alpha} \int d\mathbf{r}\rho_{\alpha}(\mathbf{r})\left[\ln\left(\rho_{\alpha}\left(\mathbf{r}\right)\right)-1\right],
\end{equation*}
and the functional derivative for ideal gas contribution is:
\begin{equation}
\frac{\delta A^{id}[\rho]}{\delta\rho_\alpha(\textbf{r})} = k_BT\ln \rho_{\alpha}(\mathbf{r}).
\label{eq:id_derivative}
\end{equation}

We use fundamental measure theory to approximate the reference hard-sphere fluid free energy functional $A^{hs}[\rho]$. Fundamental measure theory constructs the free energy density functional by weighting singlet fluid density using geometric properties of spheres. These weighted densities $n_{i}[\rho(\mathbf{r})]$ are surface averaged, volume averaged, mean curvature averaged and Gaussian curvature averaged densities. The free energy functional is\cite{roth2010fundamental}:

\begin{equation}
	A^{hs}[\rho] = k_BT \int d\mathbf{r} \Phi[n_\alpha(\mathbf{r})],
	\label{eq:hs_free_energy}
\end{equation}
and the hard sphere free energy density $ \Phi[n_\alpha(\mathbf{r})]$ is given by
\begin{align}
\Phi(\mathbf{r})&=-n_0(\mathbf{r})\ln(1-n_3(\mathbf{r}))\nonumber\\
&+\frac{n_1(\mathbf{r})n_2(\mathbf{r})-\mathbf{n}_{v1}(\mathbf{r})\cdot\mathbf{n}_{v2}(\mathbf{r})}{1-n_3(\mathbf{r})}\nonumber\\ 
&+\frac{n_2(\mathbf{r})^3-3n_2(\mathbf{r})\mathbf{n}_{v2}(\mathbf{r})\cdot\mathbf{n}_{v2}(\mathbf{r})}{24\pi(1-n_3(\mathbf{r}))^2}.
\label{eq:FMT1}
\end{align}

The calculations of the weighted densities and the functional derivatives for the hard sphere contribution require a few special treatment for cylindrical geometry. A detailed implementation is provided in Appendix \ref{derivatives} of this work.


The dispersion contribution is calculated by applying the WDA approach to the PC-SAFT~\cite{gross2001perturbed} bulk free energy. The WDA approach is similar to the method introduced by Sauer and Gross~\cite{sauer2017classical}. This method has been shown to provide an accurate prediction for interfacial tension of mixtures and the inhomogeneous free energy density reduces to the free energy of PC-SAFT in the bulk limit. The key part of WDA approach is the weighted density field:
\begin{align}
\bar{\rho}_{\alpha}(\mathbf{r}) &= \frac{1}{\frac{4}{3}\pi d_\alpha^3}\int \rho_{\alpha}(\mathbf{r}')\Theta(d_\alpha - |\mathbf{r} - \mathbf{r}'|)d\mathbf{r}'\nonumber\\&=\rho_{\alpha}(\mathbf{r}) * \omega_{\alpha}^{avg}(|\mathbf{r}-\mathbf{r}'|),
\label{eq:weighted4}
\end{align}
where $\Theta(r)$ denotes the Heaviside step function, $\ast$ denotes spatial convolution, and $d_\alpha$ is the segment diameter for component $\alpha$. The body average weighting function is:
\begin{equation*}
\omega_{\alpha}^{avg}(|\mathbf{r}-\mathbf{r}'|) = \frac{3}{4\pi d_\alpha^3}\Theta(d_\alpha - |\mathbf{r} - \mathbf{r}'|).
\end{equation*}
$d_\alpha$ is the temperature dependent hard sphere diameter of component $\alpha$ as used in PC-SAFT\cite{gross2001perturbed}:
\begin{equation*}
	d_\alpha = \sigma_{\alpha}\left[1-0.12\exp\left(-3\frac{\epsilon_{\alpha}}{k_B T}\right)\right]
\end{equation*}

From the coarse-grained weighted density field, the inhomogeneous PC-SAFT dispersion Helmholtz free energy functional can be determined by the following:

\begin{equation}
A^{disp}[\rho] = k_{B}T\int \bar{\rho}(\mathbf{r})\tilde{a}^{disp}(\bar{\rho}_{\alpha}(\mathbf{r}))d\mathbf{r}.
\label{eq:WDA}
\end{equation}
where $\bar{\rho}_{\alpha}(\mathbf{r})$ is the weighted density distribution given in (\ref{eq:weighted4}). Here, $\tilde{a}^{disp}(\bar{\rho}_{\alpha}(\mathbf{r}'))$ denotes the PC-SAFT dispersion free energy density calculated with the weighted density field:
\begin{align*}
\tilde{a}^{disp}(\bar{\rho}_{\alpha}(\mathbf{r})) = &-2\pi\bar{\rho}(\mathbf{r})I_1(\bar{\eta}, m)\overline{m^2\epsilon\sigma^3} 
\\&- \pi\bar{\rho}(\mathbf{r})\bar{m}C_1 I_2(\bar{\eta}, m)\overline{m^2\epsilon^2\sigma^3}.
\end{align*}

The total weighted density for mixture is a sum of all the individual partial densities:
\begin{equation*}
\overline{\rho}(\mathbf{r}) = \sum_{\alpha} \overline{\rho}_{\alpha}(\mathbf{r}),
\end{equation*}
and the mean PC-SAFT segment number is:
\begin{equation*}
\overline{m} = \sum_{\alpha} \overline x_\alpha m_\alpha.
\end{equation*}
The other terms in the free energy density $\tilde{a}^{disp}(\bar{\rho}_{\alpha}(\mathbf{r}))$ are given by:
\begin{gather*}
\overline{x}_\alpha = \overline{\rho}_{\alpha}(\mathbf{r})/\overline{\rho}(\mathbf{r}),
\\
\overline{\eta}(\mathbf{r}) = \frac{\pi}{6}\sum_{\alpha} \overline{\rho}_{\alpha}(\mathbf{r}) m_\alpha\sigma_\alpha^3,
\\
\overline{m^2\epsilon^n\sigma^3} = \sum_\alpha \sum_\beta \overline x_\alpha \overline x_\beta m_\alpha m_\beta \left(\frac{\epsilon_{\alpha\beta}}{kT}\right)\sigma^3_{\alpha\beta},
\\
C_1 = \left[1 + \overline{m}\frac{8\overline{\eta} - 8\overline{\eta}^2}{(1-\overline{\eta})^2} + (1-\overline{m})\frac{20\overline{\eta}-27\overline{\eta}^2+12\overline{\eta}^3-2\overline{\eta}^4}{(1-\overline{\eta})^2(2-\overline{\eta})^2}  \right]^{-1}.
\end{gather*}
The universal integral $I_1$ and $I_2$, and the molecular parameters $\sigma_\alpha$, $\epsilon_{\alpha}$, and $m_\alpha$ are the same as the ones used by PC-SAFT equation of state\cite{gross2001perturbed}. 

The calculation of the functional derivative due to dispersion contribution is obtained by differentiating the Helmholtz free energy given in (\ref{eq:WDA}):
\begin{equation}
\frac{\delta A^{disp}[\rho]/k_BT}{\delta \rho_\alpha(\mathbf{r})} = \frac{\partial \left[\bar{\rho}(\mathbf{r})\tilde{a}^{disp}(\bar{\rho}_{\alpha}(\mathbf{r}))\right]}{\partial \bar{\rho}_\alpha(\mathbf{r})} * \omega_{\alpha}^{avg}(|\mathbf{r}-\mathbf{r}'|).
\label{eq:disp_drivative}
\end{equation}

The associating free energy functional was first developed by Chapman and Segura\cite{segura1995associating,segura1998comparison} for inhomogeneous fluid. It is further extended to associating polymer by Bymaster and Chapman\cite{bymaster2010saft}.

\begin{equation}
A^{ass}[\rho] = k_BT\int d\textbf{r}'\sum_{\alpha}\rho_\alpha(\textbf{r}')\sum_{A_\alpha}\left(\ln\chi_{A_\alpha}(\textbf{r}')-\frac{\chi_{A_\alpha}(\textbf{r}')}{2}+\frac{1}{2}\right)
\label{eq:ass_functional}
\end{equation}

where $\chi_{A_\alpha}(\textbf{r})$ is the fraction of segment $\alpha$ at position $\textbf{r}$ not bonded at site $A$. The inhomogeneous unbonded fraction involves solving integral equations in real space in order to provide accurate degree of association:
\begin{equation}\label{eq:ass}
\chi_{A_\alpha}(\textbf{r})=\left[1+\int d\textbf{r}'\sum_{\alpha}\rho_\alpha(\textbf{r}')\sum_{B_\beta}\chi_{B_\beta}(\textbf{r}')\Delta^{{A_\alpha}{B_\beta}}(\textbf{r},\textbf{r}')\right]^{-1}
\end{equation}

The association strength is similar to its original form in SAFT except that it contains an inhomogeneous cavity function:

\begin{equation}
\Delta^{{A_\alpha B_\beta}}(\textbf{r},\textbf{r}')=\kappa[\exp(\beta\epsilon^{ass}_{A_\alpha B_\beta})-1]y^{\alpha \beta}(\textbf{r},\textbf{r}')
\end{equation}
where $\kappa$ is a constant that inlcudes the bonding volume and orientation constraints, and $\epsilon^{ass}_{A_\alpha B_\beta}$ is the association energy between site $A$ and site $B$ from species $\beta$ and $\alpha$, respectively. We approximate the inhomogeneous correlation function $y^{ \alpha\beta}(\textbf{r},\textbf{r}')$ for the reference hard sphere fluid between segment $\alpha$ and $\beta$ as the geometric average of  bulk radial distribution functions at contact evaluated at average density $\bar{\rho}_{\alpha}(\textbf{r})$\cite{tripathi2005microstructure}:

\begin{equation}
y^{\alpha \beta}(\textbf{r},\textbf{r}') = [g_{\alpha \beta}(\bar{\rho}(\textbf{r}),\sigma_{\alpha \beta})]^{1/2}[g_{\alpha \beta}(\bar{\rho}(\textbf{r}'),\sigma_{\alpha \beta})]^{1/2}
\end{equation}

\begin{align}
\frac{\delta A^{ass}/k_BT}{\delta\rho_{\alpha}(\textbf{r})} = & \sum_{A_\alpha}\ln\chi_{A_\alpha}(\textbf{r}) \nonumber\\&- \frac{1}{2}\int d\textbf{r}'\sum_{\alpha}\sum_{\beta}\rho_{\alpha}(\textbf{r}')\sum_{B_\beta}(1-\chi_{B_\beta}(\textbf{r}'))\left[ \frac{\delta\ln y^{\alpha \beta}(\textbf{r}')}{\delta\rho_{\alpha}(\textbf{r})}\right] 
\label{eq:ass_derivative}
\end{align}

The chain free energy functional derivative is derived from the association free energy functional at the limit of complete association\cite{jain2007modified}. 

\begin{align}
\frac{\delta A^{ch}/k_BT}{\delta\rho_{\alpha}(\textbf{r})} = & \sum_{A_\alpha}\ln\chi_{A_\alpha}(\textbf{r}) \nonumber\\&- \frac{1}{2}\int d\textbf{r}'\sum_{\alpha}\sum_{\alpha'}\rho_{\alpha}(\textbf{r}')\left[ \frac{\delta\ln y^{\alpha \alpha'}(\textbf{r}')}{\delta\rho_{\alpha}(\textbf{r})}\right]
\label{eq:ch_derivative}
\end{align}
here $\alpha'$ refers to all the segments that are bonded to segment $\alpha$.

By substituting functional derivatives  (\ref{eq:id_derivative})(\ref{eq:hs_derivative})(\ref{eq:disp_drivative})(\ref{eq:ass_derivative})(\ref{eq:ch_derivative}) into Euler--Lagrange equation~(\ref{eq:euler-lagrange}), equilibrium density distributions can be solved by:
\begin{equation}
\label{eq:EL2}
\rho_{\alpha}(\textbf{r}) = \exp(\beta\mu_M)\exp[D_\alpha(\textbf{r})- \beta V^{ext}(\textbf{r})]I_{1,\alpha}(\textbf{r})I_{2,\alpha}(\textbf{r})
\end{equation}
where
\begin{align}
D_\alpha(\textbf{r}) = &-\frac{\delta A^{hs}/k_BT}{\delta\rho_{\alpha}(\textbf{r})} \nonumber\\&-\frac{\delta A^{att}/k_BT}{\delta\rho_{\alpha}(\textbf{r})} \nonumber\\&-\frac{\delta A^{ass}/k_BT}{\delta\rho_{\alpha}(\textbf{r})} \nonumber\\&+\frac{1}{2}\int d\textbf{r}'\sum_{\alpha}\sum_{\alpha'}\rho_{\alpha}(\textbf{r}')\left[ \frac{\delta\ln y^{\alpha \alpha'}(\textbf{r}')}{\delta\rho_{\alpha}(\textbf{r})}\right],
\end{align}
and $\mu_M$ is the bulk molecular chemical potential given by a summation of bulk segment chemical potential $\mu_\alpha$ of a chain:
\begin{equation}
\mu_M = \sum_{\alpha}^{m}\mu_\alpha.
\end{equation}
Here $m$ is the total number of segments. Two recursive integrals $I_{1,\alpha}$ and $I_{2,\alpha}$ are essential to model chain molecules. Information of covalent bonding is recursively shared among all the segments of polymers by a recurrence equation. Their explicit forms are:
\begin{equation}
I_{1,\alpha}(\textbf{r}) = \int d\textbf{r}' I_{1,\alpha-1}(\textbf{r}')\exp[D_{\alpha-1}(\textbf{r}')]\Delta^{\alpha-1,\alpha}(\textbf{r}', \textbf{r})
\end{equation}

\begin{equation}
I_{2,\alpha}(\textbf{r}) = \int d\textbf{r}' I_{2,\alpha+1}(\textbf{r}')\exp[D_{\alpha+1}(\textbf{r}')]\Delta^{\alpha,\alpha+1}(\textbf{r}', \textbf{r})
\end{equation}
the boundary conditions for $I_1$ and $I_2$ are $I_{1,1} = 1$ and $I_{2,m} = 1$

\section{Fast Hankel Transform Algorithm}
\label{algo}

\subsection{Fourier Transform in Cylindrical Geometry}
Solving the Euler--Lagrange equation~(\ref{eq:euler-lagrange}) requires iterative calculation of weighted densities~in (\ref{eq:weighted_1}) and (\ref{eq:weighted4}), and functional derivatives~(\ref{eq:hs_derivative})(\ref{eq:disp_drivative})(\ref{eq:ass_derivative})(\ref{eq:ch_derivative}) which are computationally expensive since convolutions are involved. A natural approach to circumvent this step is to apply the convolution theorem and the FFT algorithm. Comparing with directly solving the Euler--Lagrange equation~(\ref{eq:euler-lagrange}), utilizing the FFT algorithm usually provides a speedup from $\mathcal{O}(N^2)$ to $\mathcal{O}(N\log N)$. This method is well applied to a system with Cartesian mesh~\cite{sears2003new}. 

The forward and backward two-dimensional Fourier transform of a two-dimensional function are:

\begin{gather}
g(\mathbf{k}) = \int f(\mathbf{r})\exp(i\mathbf{k}\cdot \mathbf{r}) d\mathbf{r},
\label{eq:transform}
\\
f(\mathbf{r}) = \frac{1}{(2\pi)^2} \int g(\mathbf{k})\exp(-i\mathbf{k}\cdot\mathbf{r})d\mathbf{k},
\label{eq:inverse}
\end{gather}
where $f(\mathbf{r}) = f(x,y)$ and $g(\mathbf{k}) = g(k_x,k_y)$. If the function $f(\mathbf{r})$ is a circularly symmetric function, i.e. $f(\mathbf{r}) = f(r)$ with $r = \sqrt{x^2+y^2}$, its Fourier transform is also a circularly symmetric function $g(\mathbf{k}) = g(k)$ with $k = \sqrt{k_x^2+k_y^2}$. More precisely, the forward and backward Fourier transforms of a circular symmetric function are Hankel transforms of zeroth order:

\begin{gather}
g(k) = 2\pi \int_{0}^{\infty} rf(r)J_0(2\pi rk)dr,
\label{eq:general_1}
\\
{f}_{}(r) =  \frac{1}{2\pi} \int_{0}^{\infty} kg(k)J_0(2\pi rk)dk.
\label{eq:general_2}
\end{gather}

For inhomogeneous fluid in cylindrical geometry, the density distribution $\rho_{\alpha}(\mathbf{r})$ reduces to $\rho_{\alpha}(r)$. (\ref{eq:transform}) and (\ref{eq:inverse}) reduce to Hankel transform:

\begin{gather}
\tilde{\rho}_{\alpha}(k) = 2\pi \int_{0}^{\infty} r\rho_{\alpha}(r)J_0(kr)dr,
\label{eq:h_transform}
\\
{\rho}_{\alpha}(r) = \frac{1}{2\pi} \int_{0}^{\infty} k\tilde{\rho}_{\alpha}(k)J_0(kr)dk.
\label{eq:h_inverse}
\end{gather}
where $\tilde{\rho}_{\alpha}(k)$ is the Hankel transform of density distribution ${\rho}_{\alpha}(r)$ in cylindrical geometry. $J_0$ is zeroth-order Bessel function of the first kind. Now (\ref{eq:h_transform}) and (\ref{eq:h_inverse}) are Hankel transforms of zeroth-order. (\ref{eq:h_transform}) and (\ref{eq:h_inverse}) have to be numerically calculated. Lado~\cite{lado1971numerical} introduced the finite discretized Hankel transform by using the orthogonality between the zeroth-order and first-order Bessel functions\cite{gonzalez2006density}:

\begin{equation}
\label{l_transform}
\tilde{\rho}_{\alpha}(k_j) = \sum_{i=1}^{N}\frac{4\pi J_0(k_jr_i)}{k_N^2J_1^2(z_i)}\rho_{\alpha}(r_i),
\end{equation}
and its backward transform is:

\begin{equation}
\label{l_inverse}
{\rho}_{\alpha}(r_i) = \sum_{j=1}^{N}\frac{ J_0(k_jr_i)}{\pi r_N^2J_1^2(z_j)}\tilde{\rho}_{\alpha}(k_j).
\end{equation}
The $r_i$ and $k_j$ are discretized real space grid and Fourier space grid, and $z_i$ is the $i^\text{th}$ zero of $J_0$. (\ref{l_transform}) and (\ref{l_inverse}) are matrix-vector products and the computation scales as $\mathcal{O}(N^2)$ with $N$ as number of mesh grid. Lado's method~\cite{lado1971numerical} therefore does not provide an efficient $\mathcal{O}(N\log N)$ time scaling.

\subsection{Fast Hankel Transform}
Additional speedup in computing functional derivatives can be achieved by exploiting a transformation of variables~\cite{magni1992high}. The space variable $r$ and frequency variable $k$ are first normalized by a spatial domain cut-off $b$ and a frequency cut-off $\lambda$. The function $f(r)$ vanishes when $r$ is beyond $b$, more precisely $f(r)=0$ if $r\ge b$. The space cut-off and the frequency cut-off are related by the space-bandwidth product $\gamma=\lambda b$. 

By substituting with normalized variables $x=r/b$ and $y=k/\lambda$, the truncated and normalized Hankel transforms (\ref{eq:general_1}) and 
(\ref{eq:general_2}) are :

\begin{gather}
g(y) = 2\pi \gamma \frac{b}{\lambda} \int_{0}^{1} xf(x)J_0(2\pi\gamma xy)dx,
\label{eq:new_transform1}
\\
{f}_{}(x) = \frac{1}{2\pi} \gamma \frac{\lambda}{b}\int_{0}^{1} yg(y)J_0(2\pi\gamma xy)dy.
\label{eq:new_transform2}
\end{gather}

To numerically compute (\ref{eq:new_transform1}) and (\ref{eq:new_transform2}), the domain i.e. $0\le x \le 1$ needs to be further divided into $N$ sub-intervals by selecting the partition points $\chi_i$, where $0 = \chi_0 < \chi_1 < ... < \chi_N = 1$. The original function $f(x)$ is evaluated at $x_i$ ($i = 0,1,...,N-1$), where $\chi_i \leq x_i \leq \chi_{i+1} $. $f(x_N) = 0$ is set as a boundary condition. The sub-intervals are defined in this way such that there is exactly one point $x_i$ per sub-interval. Because of the partition, it is reasonable to assume that $f(x) \approx f(x_i)$ for $x$ within the interval i.e. $\chi_i < x_i < \chi_{i+1}$. The divided integral of (\ref{eq:new_transform1}) and (\ref{eq:new_transform2}) for each sub-interval can be expressed as:

\begin{align}
\label{analytical}
&\int_{\chi_i}^{\chi_{i+1}} xf(x)J_0(2\pi\gamma yx)dx \nonumber \\&= \frac{f(x_i)}{2\pi\gamma y} [J_1(2\pi\gamma y\chi_{i+1})\chi_{i+1} - J_1(2\pi\gamma y\chi_i)\chi_i],
\end{align}
where the second part of (\ref{analytical}) is an analytical result if we apply the approximation $f(x) \approx f(x_i)$ for $\chi_i < x_i < \chi_{i+1}$ and realize the fact $\int xJ_0(x) dx = xJ_1(x) + C$. Summing over all the intervals and considering two boundary conditions $J_1(0) = 0$ and $f(x_N) = 0$, (\ref{eq:new_transform1}) can be numerically evaluated:

\begin{equation}
\label{J1}
g(y) = \frac{b}{\lambda y} \sum_{i=0}^{N-1} [f(x_i)-f(x_{i+1})]J_1(2\pi\gamma y\chi_{i+1})\chi_{i+1}.
\end{equation}

The discretization process is based on exponential transformation. A natural approach to set the partition point $\chi_i$ is: 
\begin{align}
\chi_i = 0 \quad &\text{for} \quad i = 0\nonumber\\
\chi_i = e^{\alpha(i-N)} \quad &\text{for} \quad i = 1,2,...,N
\label{eq:chi_trans}
\end{align}
such that $0 = \chi_0 < \chi_1 < ... < \chi_N = 1$ is satisfied. 

The grids for $x_i$ and $y_j$ are also non-equidistant:
\begin{gather}
	\label{eq:x_trans}
	x_i = x_0 e^{\alpha i} \quad \text{for} \quad i = 0,1,2,...,N-1,\\
	y_j = x_0 e^{\alpha j} \quad \text{for} \quad j = 0,1,2,...,N-1.
	\label{eq:y_trans}
\end{gather}

The reason for applying the exponential transformation is evident by analyzing the general Hankel transforms (\ref{eq:general_1}) and (\ref{eq:general_2}). If we treat the continuous variables $r$ and $k$ in (\ref{eq:general_1}) and (\ref{eq:general_2}) as $r=r_0e^{\alpha u}$ and $k=r_0e^{\alpha v}$, and substitute them into (\ref{eq:general_1}) and (\ref{eq:general_2}). We end up with the following cross-correlations that can be efficiently evaluated by FFT algorithm:

\begin{align*}
	\hat{g}(v) = 2\pi\int_{-\infty}^{\infty}\hat{f}(u)\hat{j}(u+v)du,\\
	\hat{f}(u) =
	\frac{1}{2\pi}\int_{-\infty}^{\infty}\hat{g}(v)\hat{j}(u+v)dv\\
\end{align*}
where $\hat{f}(u) = rf(r)$, $\hat{g}(v) = kg(k)$, and $\hat{j}(u+v)=\alpha rkJ_0(2\pi rk)$.

If we apply the discretization (\ref{eq:x_trans}-\ref{eq:y_trans}) to (\ref{J1}), we obtain $\tilde{g}(y_j)$, a discretized $g(y)$:
\begin{align}
\tilde{g}(y_j) =\frac{b}{\lambda y_j} \sum_{i=0}^{N-1} &k_i[f(x_i)-f(x_{i+1})] e^{\alpha(i+1-N)}\nonumber\\&J_1(2\pi\gamma x_0 e^{\alpha(i+j-1-N)}).
\label{eq:cross-corr}
\end{align}

Now (\ref{eq:cross-corr}) can also be treated as a discrete cross-correlation:

\begin{gather}
	\tilde{g}(y_j) = \frac{b}{\lambda y_j} \sum_{i=0}^{N-1} \tilde{\phi}_i \tilde{j}_{i+j}
	\label{eq:dis_fft}
\end{gather}
where we define:

\begin{equation*}
\tilde{\phi}_i = \begin{cases}
k_i[f(x_i) - f(x_{i+1})]e^{\alpha(i+1-N)} \quad &\text{for} \quad i = 0,...,N-1,\\
0 \quad &\text{for} \quad i = N,...,2N-1.
\end{cases}
\end{equation*}
and the kernel:
\begin{equation*}
\tilde{j}_{i+j} = J_1(2\pi\gamma x_0 e^{\alpha(i+j-1-N)}).
\end{equation*}

We can apply the efficient FFT algorithm to calculate (\ref{eq:dis_fft}):
\begin{equation*}
\tilde{g}(y_j) = \frac{b}{\lambda y_j} \mathcal{F}\{\mathcal{F}\{\tilde{\phi}_i\}\mathcal{F}^{-1}\{\tilde{j}_{i+j}\}\},
\end{equation*}
where $\mathcal{F}$ is forward Fourier transform and $\mathcal{F}^{-1}$ is inverse Fourier transform.

It is worthwhile to note that $\alpha$ and $x_0$ in (\ref{eq:cross-corr}) are free parameters to be determined. To ensure that every $x_i$ lies in between $\chi_{i}$ and $\chi_{i+1}$ for $i = 1,2,...,N-1$, an inequality needs to be satisfied:

\begin{gather}
	e^{\alpha(i-N)} < x_0 e^{\alpha i} < e^{\alpha(i+1-N)}\nonumber\\
	e^{-\alpha N} < x_0 < e^{\alpha(1-N)}.
	\label{eq:inequality}
\end{gather}

(\ref{eq:inequality}) suggests the average of its upper bound and lower bounds is a good choice for $x_0$ i.e. $x_0 = [e^{-\alpha N}+ e^{\alpha(1-N)}]/2$. Because $x_0$ does not lie in between $\chi_{i}$ and $\chi_{i+1}$, $f(x) \approx f(x_0)$ is an erroneous approximation for $\chi_0 < x < \chi_{1}$. Hence the factor $k_0$ is an end correction factor\cite{agrawal1981end} to correct the approximation for the end sub-interval including origin. The correction factor $k_i$ is:
\begin{equation}
k_i = \begin{cases}
1 \quad &\text{for} \quad i \ne 0,
\\
\frac{2 e^{\alpha}+ e^{2\alpha}}{[1+ e^{\alpha}]^2[1-e^{2\alpha}]} \quad &\text{for} \quad i = 0.
\end{cases}
\label{eq:correction_factor} 
\end{equation}

The derivation of the end correction factor arises from substituting $f(x_0)$ with $f(x_0^\dagger)$, where $x_0^\dagger = (\chi_0 + \chi_1)/2 = \chi_1 /2.$ Then we can use $f(x_0)$ and $f(x_1)$ to interpolate $f(x_0^\dagger)$ by quadratic parabola $f(x) = a_2x^2 + a_0$. It has no first order term since we assume the first order derivative of the quadratic parabola vanishes at origin. We arrive at the following relation:

\begin{align}
	f(x_0^\dagger) - f(x_1) = & \frac{[f(x_0)-f(x_1)]}{x_0^2 - x_1^ 2}\left({x_0^\dagger}^2 - x_1^2 \right)\nonumber\\ =&[f(x_0) - f(x_1)]\frac{[\frac{1}{4}e^{2\alpha(1-N)} - x_0^2e^{2\alpha}]}{x_0^2[e^{2\alpha} -1]}
	\label{eq:correct}
\end{align} 
Further simplifying (\ref{eq:correct}), we obtain the correction factor that $f(x_0^\dagger) - f(x_1) = k_0 [f(x_0) - f(x_1)]$.

It comes to authors' attention that Bo\c tan et al\cite{boctan2009hard} previously introduced a fast Hankel transform for density functional theory. Bo\c tan's treatment is similar to the original QFHT by Siegman\cite{siegman1977quasi} without end correction term. The difference from this work is that we approximate the function as constant over each sub-interval which leads to different kernel basis we used. The time scaling of the implementation is similar to this work. Bo\c tan's treatment also ignores the end correction term i.e. it does not include the interval $0 < x < x_{0}$. The missing end correction term can possibly lead to divergence at origin which we will show in next section. A more comprehensive discussion on end correction term can be seen in the reference~\cite{agrawal1981end}.

\subsection{Choice of Parameters for Fast Hankel Transform}

In order to apply the new algorithm, a few parameters have to be determined, such as the position of the first mesh $x_0$, the maximum frequency cut-off $\lambda$, the maximum space cut-off $b$, the number of mesh points $N$, and free transformation parameters $\alpha$. The accuracy of the algorithm depends on the bandwidth $\gamma$, which is the product of $\lambda$ and $b$. The choice of space cut-off $b$ can be determined from a priori knowledge about the system domain such as cylindrical pore radius $R$. 

The choice of frequency cut-off $\lambda$ depends on the nature of function $f(x)$ itself, which is the density distribution $\rho(r)$ in this study. In our study, it is found that $\lambda$ increases with system domain size. An effective $\lambda$ from 30 to 50 is used in this study when the dimensionless system domain size varies from R*=3 to R*=5. The number of grid points $N$ is chosen to be from $2^{9}$ to $2^{11}$ for varying system domain. A choice for $x_0$ has been given the previously. Another free parameter $\alpha$ can be determined by setting the same width for the first and the last sub-intervals i.e. $\chi_{N} - \chi_{N-1} = \chi_{1} - \chi_{0}$\cite{magni1992high}. This gives a good choice for $\alpha$:
\begin{gather*}
\alpha = -\ln[1-e^{\alpha}]/(N-1).
\end{gather*}

Though it is desirable to have a higher frequency cut-off $\lambda$ to capture the rapid oscillation of fluid distribution, it leads to insufficient sampling at high frequency. More grid points $N$ always increases sampling effectiveness, but it is computationally expensive. A good criterion to balance those two parameters is to enforce the least number of cycles between two zeros of the Bessel kernel $J_0$. The difference between two zeros of Bessel function is approximately equal $\pi$. If there is at least one sampling between two zeros, then a relation based on (\ref{eq:new_transform1}) and (\ref{eq:new_transform2}) is:

\begin{equation}
2\pi \gamma x_0 \left[e^{\alpha(N-1)} - e^{\alpha(N-2)}\right] < \pi.
\end{equation}

This inequality enforces that at least one Bessel zero is sampled over the last two sampling points where it has the largest grid spacing.

\section{Analysis of the Algorithm}
In this section, we first compare the density distributions for hard-sphere fluids in cylindrical geometry by DFT solved by the proposed algorithm with the grand canonical Monte Carlo simulation (GCMC) by Mlijevsky\cite{malijevsky2007fundamental}. A comprehensive analysis of convergence, performance and stability of the algorithm is followed.

\subsection{Verification of the Algorithm by Binary Hard-spheres in Hard Cylindrical Pores}
\begin{figure}[!t]
	\centering
	\includegraphics[scale=0.6]{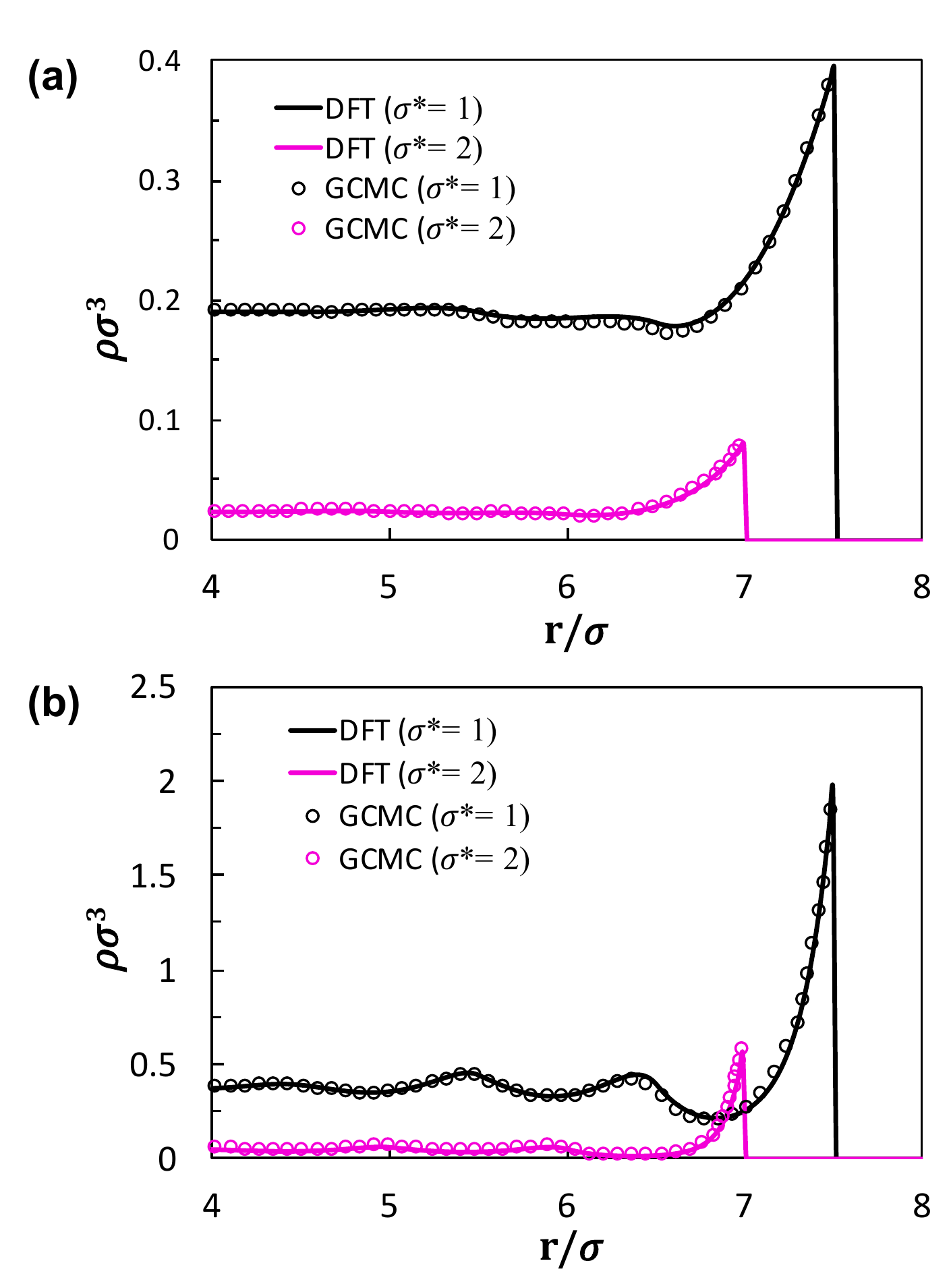}
	\caption{A binary mixture of hard-spheres of size ratio 2:1 in a hard cylindrical nano-pore at overall bulk packing fraction (a) $\eta$=$0.3$ and (b) $\eta$=$0.4$. Solid lines are results from this work. Symbols are digitized molecular simulation results of Malijevsky~\cite{malijevsky2007fundamental}.}
	\label{fig:hs}
\end{figure}

Fig. \ref{fig:hs} shows density distribution of a binary mixture of hard-spheres confined in hard cylindrical pores. The external potential of a hard cylindrical pore is given by:

\begin{equation*}
V^{ext}_{\alpha}(r) = \begin{cases}
\infty, \quad r > R-\frac{\sigma_\alpha}{2},\\
0, \quad r < R-\frac{\sigma_\alpha}{2}.
\end{cases}
\end{equation*}

Hard cylindrical pores with pore radii $R/\sigma_1=8$ and overall bulk packing fractions $\eta$=$0.3$ and $\eta$=$0.4$ are chosen. The packing fraction is defined by $\eta$=$\sum_{\alpha}\pi\rho_{\alpha}\sigma_\alpha^3/6$. The size ratio of large to small hard-spheres is set as $\sigma_2/\sigma_1=2$. Good agreement between molecular simulation and density functional theory solved by the new method is found from low to high packing densities in Fig. \ref{fig:hs} (a) and Fig. \ref{fig:hs} (b). At low packing fraction $\eta$=$0.3$, both theory and simulation show there is little fluid structure except at wall for both small and large particles. At high packing fraction $\eta$=$0.4$, the small particle has a three-layered strongly oscillatory fluid structure and two layers of large particles  are adsorbed on the hard cylindrical wall.

\subsection{Convergence, Performance, and Stability}

Since we are interested in applying the new algorithm to density functional theory, here we discuss the convergence, performance, and stability of the new algorithm. For simplicity and consideration of time, the analysis here is for hard-sphere fluids in hard cylindrical pores of pore radii R$^*_{max}$ at a reduced bulk density $\rho^*_B$=0.8 at different domain sizes. Here $*$ means dimensionless unit. The mixing fraction for Picard iteration is constant 1E-2. The grid size is set to $\Delta$r$^*$=0.02 for Lado's method and direct integration method. Since the new algorithm has a non-equidistant grid, we set the last grid spacing for the new method to $\Delta$r$^*_{max}$=0.02 for comparison. 

\begin{figure}[!t]
	\centering
	\includegraphics[scale=.5]{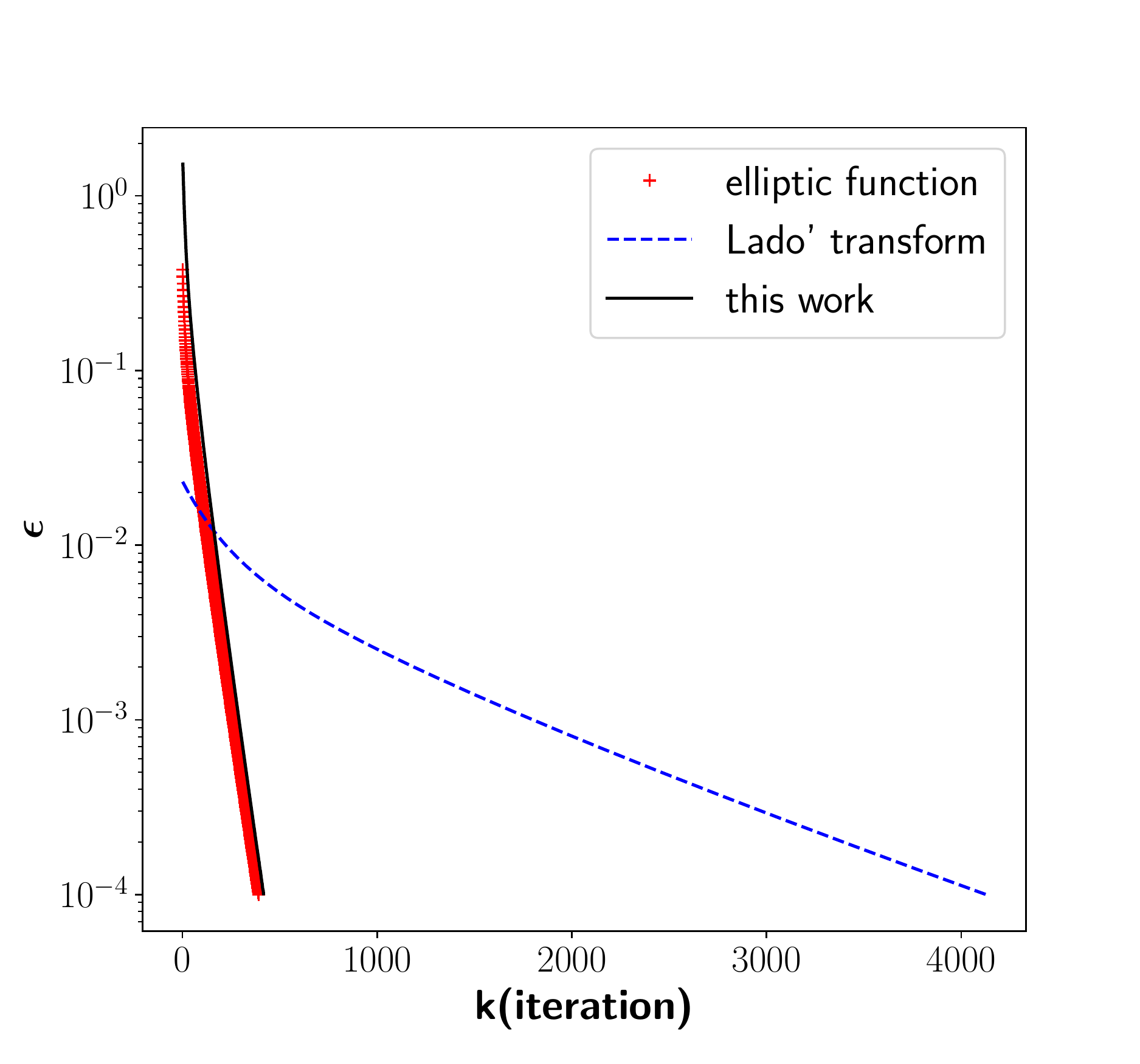}
	\caption{Residual norm changes with the number of Picard iterations for hard-spheres adsorbed in a hard cylindrical pore of radius R$^*_{max}$=8 at reduced bulk density $\rho^*_B$=0.8.}
	\label{fig:convergence}
\end{figure}

The convergence is represented by using the residual norm:
\begin{equation*}
\epsilon = \left(\int\left(\rho^*_{k+1}(\mathbf{r}^*) - \rho^*_{k}(\mathbf{r}^*)\right)^2  d\mathbf{r}^*\right)^{1/2}
\end{equation*}
where $\rho^* = \rho\sigma^3$ and $r^* = r/\sigma $. It is the square root of the total squared residual of all the density distribution from the $k^\text{th}$ to the $(k+1)^\text{th}$ iteration in solving (\ref{eq:euler-lagrange}). The convergence of the implementations by ecliptic functions, Lado's method, and this work at the state point is shown in Fig \ref{fig:convergence}. It is interesting that the convergence of the new algorithm and the method using elliptic functions are similar. They both reach a residual norm 1E-4 after 500 iterations. However, Lado's method has a slower convergence rate. It requires 4000 iterations to reduce the residual norm to 1E-4.

\begin{figure}[!t]
	\centering
	\includegraphics[scale=.5]{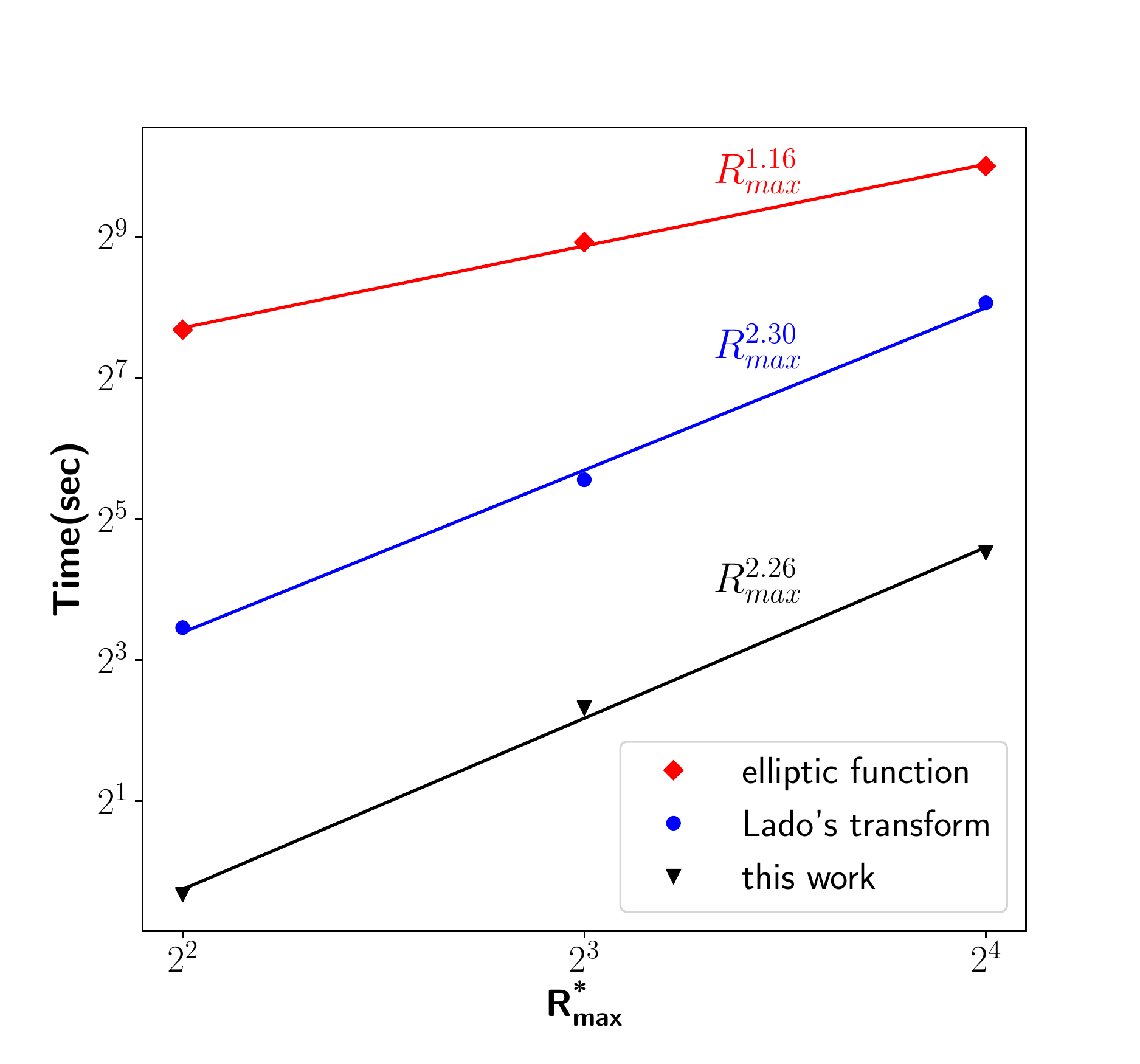}
	\caption{Time scaling for hard-sphere for hard-spheres at reduced bulk density $\rho^*_B$=0.8 adsorbed in hard cylindrical pores of varying pore sizes. The time to convergence for each algorithm is measured on a single desktop machine of Intel Core i7-6700 processor with a base frequency 3.40 GHz.}
	\label{fig:performance}
\end{figure}

Though the convergence rates vary for different implementations, we are ultimately interested in knowing the machine time to solve density functional theory across different algorithms. Here we show the scaling of time with respect to the system domain size since the grid spacing is not uniform for the new algorithm. The domain size R$^*_{max}$ ranges from 2 to 16. We compare all the three implementations under the constraint that the largest grid spacings are the same. Fig \ref{fig:performance} shows the performance of all three algorithms. The convergence of the elliptic function method is faster than the Lado's method in Fig \ref{fig:convergence} though the Lado's method is faster than the elliptic function method as shown in Fig \ref{fig:performance}. This is because the computation of digamma functions required by elliptic function method is very expensive. The new algorithm is faster than the Lado's method by another order of magnitude. Though they are both Fourier transform methods, Lado's method requires an $\mathcal{O}(N^2)$ operation as discussed previously, and FFT algorithm cannot be applied. The new algorithm in this work applies exponential transformation in (\ref{eq:x_trans}-\ref{eq:y_trans}). This turns Hankel transforms (\ref{eq:new_transform1}-\ref{eq:new_transform2}) into a cross-correlation in (\ref{eq:dis_fft}). Therefore the efficient implementation of FFT algorithm can be directly applied. The scaling of the new algorithm is not the best because of the non-equidistant grids. However, it shows an excellent speedup for a system with a small domain size in Fig~\ref{fig:performance}. This makes the algorithm especially efficient for the study of fluid adsorption in a nano-pore of zeolites which have a pore size of a few angstroms.

It is worth to mention that additional speedup in Fourier transform is achieved because the weighting functions in fundamental measure theory have closed-form solutions in Fourier space, and they can be evaluated in $\mathcal{O}(1)$ operation. Details in deriving the closed-form weighting functions can be found in Appendix \ref{transform}. This further explains why the transform methods are always faster than the elliptic function method.

\begin{figure}[!t]
	\centering
	\includegraphics[scale=.7]{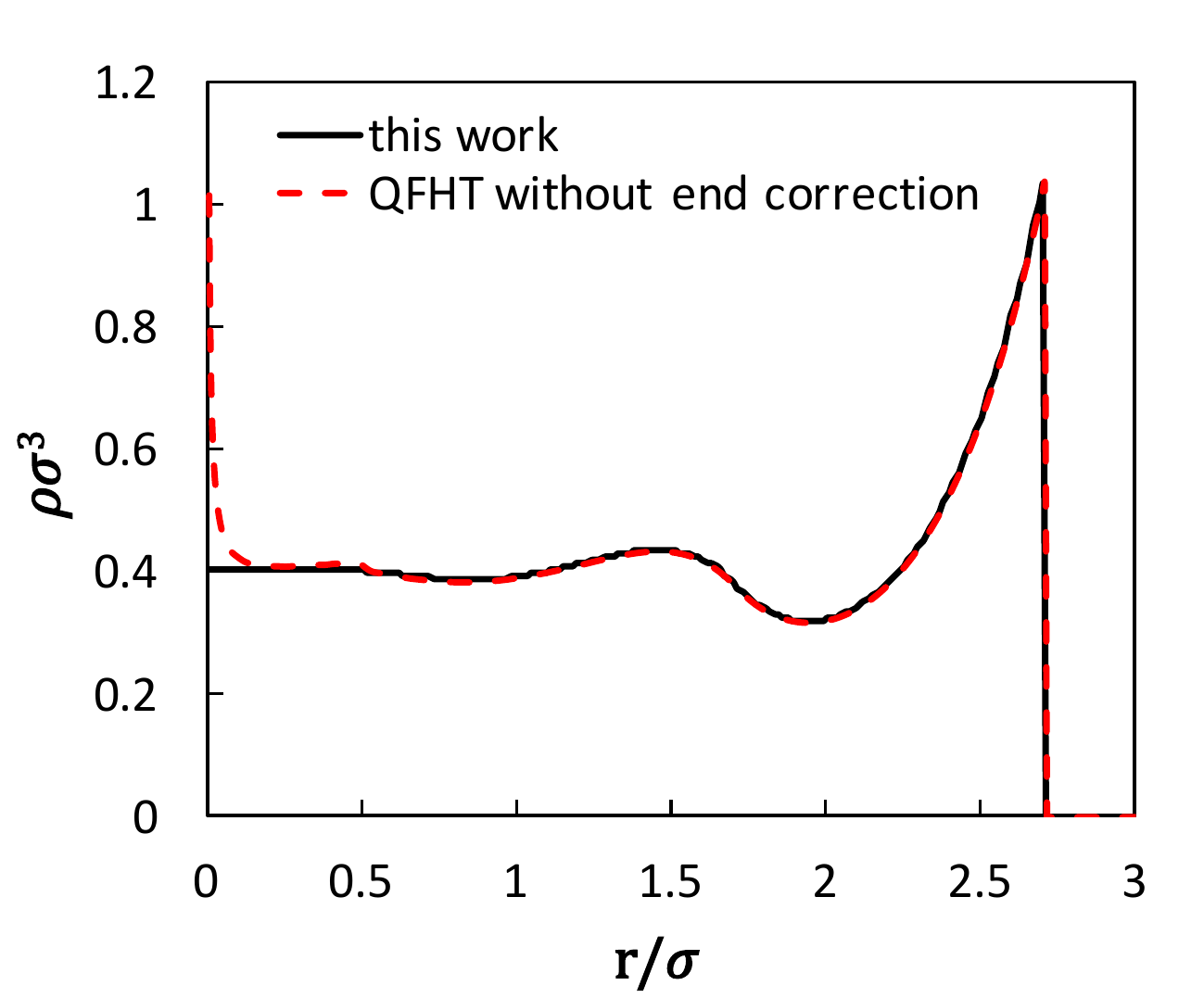}
	\caption{Solution provided by our method (solid line) and solution provide by QFHT(dashed line) for hard-sphere fluid in a hard cylindrical pore with R*=3 and bulk density $\rho_B^*$=0.4 with a maximum mesh size $\Delta$r$^*_1$=3E-3. QFHT does not include a correction factor at origin.}
	\label{fig:stability}
\end{figure}

The Hankel transform can be regarded as an expansion of a function by Fourier--Bessel series. Fourier transform has a Gibbs phenomenon where the function to be transformed has a discontinuity. However, it is interesting that the Fourier--Bessel expansion also exhibits a Gibbs-like phenomenon close to the origin of radial direction. Pinsky~\cite{gray1992computer} showed that the asymptotic behavior of Fourier--Bessel series has a slower rate of convergence at the origin than at surrounding points. For the zeroth order Fourier-Bessel expansion, which is the basis for the Hankel transform in this work, of $f(r)=1$, its asymptotic behavior when $k\rightarrow \infty$ and $r\neq0$ is~\cite{gray1992computer}:

\begin{equation}
\frac{\left(-1\right)^{k-1}\cos\left(\left(1+r-4kr\right)\pi /4 \right)}{\left(k-1 / 4\right)\pi \sqrt{r}} + \mathcal{O}\left(\frac{1}{k^2}\right),
\label{eq:conv1}
\end{equation}
for $k\rightarrow \infty$ and $r=0$ the asymptotic form of $f(r)=1$ is:
\begin{equation}
\frac{\left(-1\right)^{k-1}\sqrt{2}}{\sqrt{4k-1}} + \mathcal{O}\left(\frac{1}{k^{3/2}}\right).
\label{eq:conv2}
\end{equation}
the rate of convergence at the origin is $\mathcal{O}(\frac{1}{k^{3/2}})$ and its surrounding converges at a faster rate of $\mathcal{O}(\frac{1}{k^2})$. The Gibbs-like phenomenon at the origin can cause a divergence at the origin of a cylindrical pore in density functional theory. Adding the correction factor to origin in (\ref{eq:correction_factor}) can improve the stability of the solution method. A situation is demonstrated in Fig \ref{fig:stability} that the system solved by a maximum grid spacing of $\Delta$r$^*$=3E-2 without including the error correction term at the origin diverges. 

\section{Application of the Algorithm for iSAFT}
\subsection{Methane Adsorbed Under a Cylindrical Steele 10-4-3 Attractive Potential}
The phase behavior of hydrocarbons in nano-scale pores is important for oil and natural gas production from tight formations\cite{ratner2014overview} and chemical processing\cite{fechete2015nanoporous}. The study of phase behavior of hydrocarbons under reservoir conditions requires an inhomogeneous thermodynamic model for hydrocarbons and surface potential. In this section, we show that iSAFT with the recently developed cylindrical Steele 10-4-3 potential can be used to study the fluid distribution of methane molecules in graphite cylindrical nano-pores by applying the new numerical algorithm. The modeling of methane adsorption requires several parameters to characterize the physical properties of methane, graphite wall, and the interaction between methane and graphite. 


The external potential we used here is the Steele 10-4-3~\cite{steele1973physical} type of cylindrical potential provided by Siderius and Gelb~\cite{siderius2011extension}. The cylindrical LJ 9-3 potential~\cite{peterson1986fluid,zhang2004potential} and the potential by Tjatjopoulos et al.~\cite{tjatjopoulos1988molecule} have been exploited in previous studies of fluid adsorption in cylindrical pores. The potential by Tjatjopoulos has been successfully applied to model adsorption in MCM-41 type materials and single-wall carbon nanotubes. It suffers from the fact that this potential is based on a single cylindrical shell model. Therefore it does not describe a confining wall with a multi-layer structure, and it cannot reduce to the Steele 10-4-3 potential in the infinite large pore limit. The cylindrical Steele 10-4-3 potential is dependent on the radial distance to the surface and pore radius. It is given as~\cite{siderius2011extension}:

\begin{align}
\label{steel}
V^{10-4-3}(r,R) = 2\pi \rho_s \Delta\sigma_s^2\epsilon_s[&\psi_6(r, R, \sigma_s)-\psi_3(r, R, \sigma_s) \nonumber\\& - \frac{\sigma_s}{\Delta}\phi_3(r,R+\alpha\Delta,\sigma_s)],
\end{align}
where $\rho_s$, $\epsilon_s$, $\Delta$, and $\sigma_s$ are parameters of the solid wall to represent the interactions between solid atoms and fluid molecules.

The term $\phi_n(r, R, \sigma_s)$ can be calculated:

\begin{align*}
\phi_n(r, R, \sigma_s) = &\frac{4\sqrt{\pi}\Gamma(n-\frac{1}{2})}{(2n-3)\Gamma(n)} (\frac{\sigma_s}{R})^{2n-3}[1 - (\frac{r}{R})^2]^{3-2n}\nonumber\\& F[\frac{3-2n}{2},\frac{3-2n}{2};1;(\frac{r}{R})^2],
\end{align*}
and $\psi_n(r, R, \sigma_s)$ can be calculated:

\begin{align*}
\psi_n(r, R, \sigma_s) = &\frac{4\sqrt{\pi}\Gamma(n-\frac{1}{2})}{\Gamma(n)} (\frac{\sigma_s}{R})^{2n-2}[1 - (\frac{r}{R})^2]^{2-2n}\\& F[\frac{3-2n}{2},\frac{5-2n}{2};1;(\frac{r}{R})^2],
\end{align*}
where $F(a,b;c;z)$ denotes the Gaussian hypergeometric function, $\Gamma(n)$ is the Gamma function, and $R$ is the radius of the cylindrical pore. It can be shown that this cylindrical Steele potential reduces to the planar Steele potential in the limit of infinitely large pore~\cite{siderius2011extension}. The pure-component parameters for methane used in this study from PC-SAFT~\cite{gross2001perturbed} are $m=1$, $\sigma=3.7039\si{\angstrom}$, and $\epsilon/k_B=150.03$ K. The binary interaction parameters are provided in Table \ref{tbl:parameters}. The interaction parameters between methane and graphite are obtained using an Lorentz--Berthelot rule. $\alpha$ in (\ref{steel}) is set as 0.61 as empirical treatment. 

\begin{table}
	\centering
	\caption{DFT Parameters for Methane Adsorption in Graphite Carbon Nano-pore}
	\label{tbl:parameters}
	\begin{tabular}{llrlc}
		\hline
		Component & $\sigma(nm)$ & $\epsilon/k_B (K) $ & $\rho_s(nm^{-3})$ & $\Delta(nm)$ \\
		\hline
		CH$_4$ & 0.37039 & 150.03 & N.A.  & N.A.\\
		Graphite & 0.34000 & 28.00 & 114  & 0.335\\
		CH$_4$-Graphite  & 0.35520 & 64.81 & N.A.  & N.A. \\
		\hline
	\end{tabular}
\end{table}

\begin{figure}[!t]
	\centering
	\includegraphics[scale=0.6]{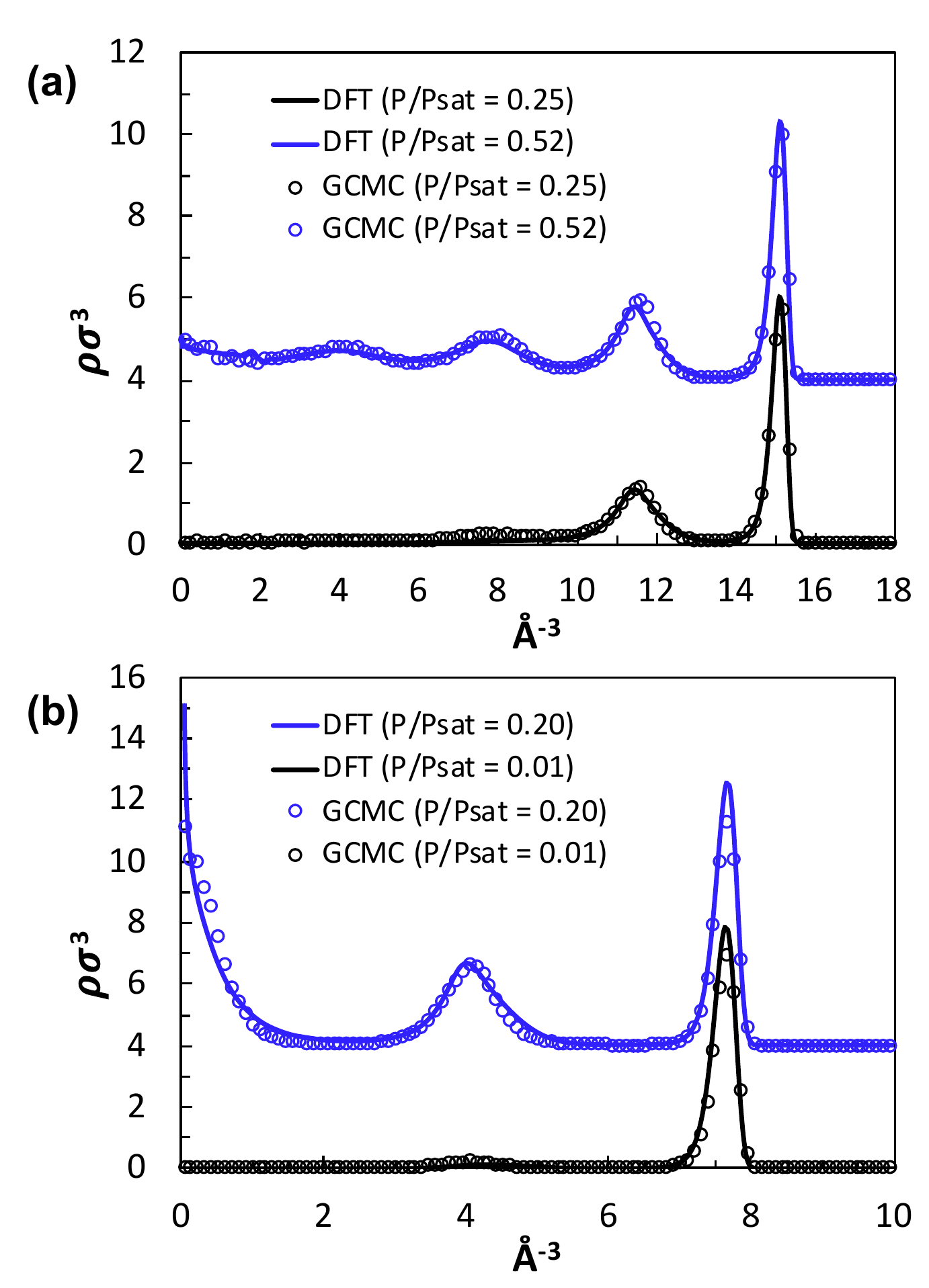}
	\caption{Density profiles of methane adsorbed in cylindrical nano-pores of radius R=18.55$\angstrom$ at 150K (a), and R=11.13$\angstrom$ at T=105K (b). Methane bulk pressures are set as 0.52Psat and 0.25Psat in (a), 0.20Psat and 0.01Psat in (b). Solid lines represent density functional theory results using the new algorithm. Markers represent our molecular simulation results.}
	\label{fig:methane}
\end{figure}

In Fig.~\ref{fig:methane} (a) we present the microscopic density profile for methane adsorbed in a large cylindrical pore at R=18.55$\angstrom$ at 150K. Two microscopic pictures of methane correspond to the profile before and after capillary condensation. Microscopic fluid structures of methane adsorbed in a small pore of R=11.13$\angstrom$ at 105K are shown in Fig.~\ref{fig:methane} (b). The packing effect is well demonstrated in Fig.~\ref{fig:methane} (b) that exceptionally high microscopic density is found at the center of the pore.

\begin{figure}[!t]
	\centering
	\includegraphics[scale=0.6]{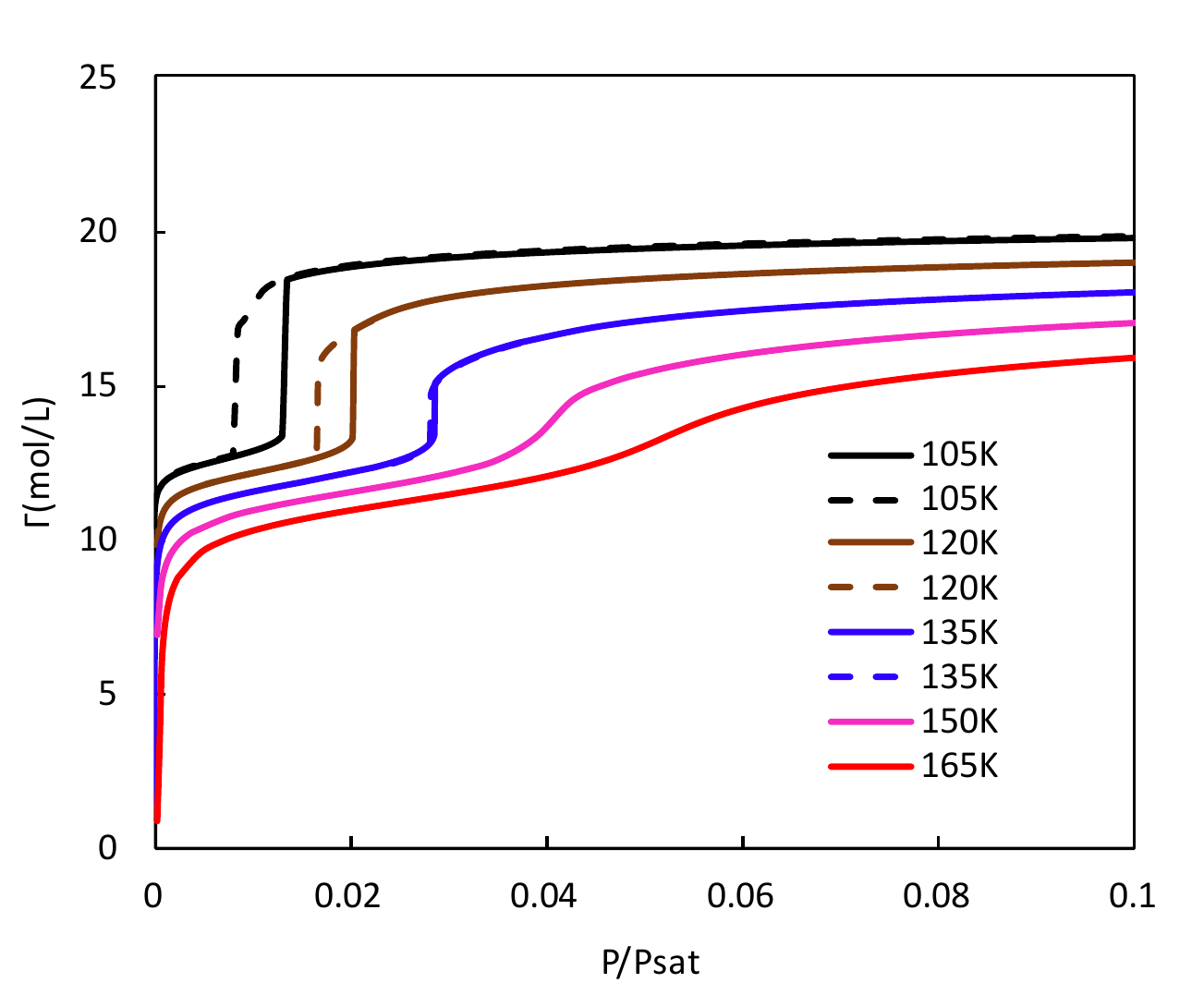}
	\caption{Methane adsorption isotherm in a cylindrical nano-pore of radius R=11.13$\angstrom$. Isotherms are calculated by DFT at temperatures T=105K, 120K, 135K, 150K, and 165K. Dashed lines and solid lines refer to desorption and adsorption, respectively.}
	\label{fig:isotherm_R3.5}
\end{figure}
\begin{figure}[!t]
	\centering
	\includegraphics[scale=0.6]{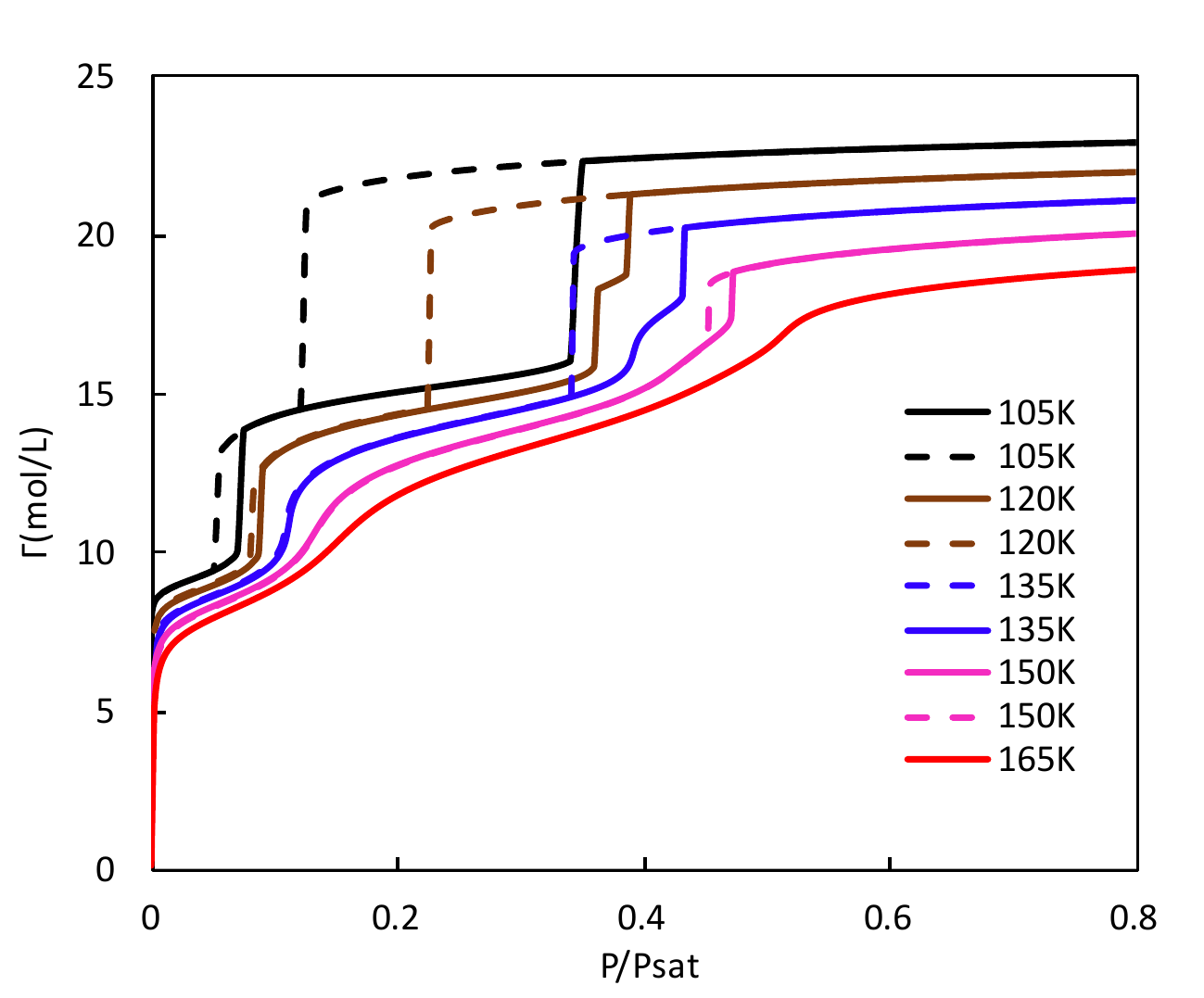}
	\caption{Methane adsorption isotherm in a cylindrical nano-pore of radius R=18.55$\angstrom$. Isotherms are calculated by DFT at temperatures T=105K, 120K, 135K, 150K, and 165K. Dashed lines and solid lines refer to desorption and adsorption, respectively.}
	\label{fig:isotherm_R5}
\end{figure}

Fig.~\ref{fig:isotherm_R3.5} and Fig.~\ref{fig:isotherm_R5} show the adsorption isotherm of methane for a cylindrical pore of radius R=11.13$\angstrom$ and R=18.55$\angstrom$. The adsorption is defined by: 

\begin{equation}
\Gamma = 2/R^2\int_{0}^{R} \rho(r)rdr,
\end{equation}
where R is the radius of a cylindrical pore. For a smaller nano-pore of radii R=11.13$\angstrom$, a single hysteresis loop is found at T=105K, T=120K, and T=135K which indicates first-order capillary condensations. Sharp layering transitions are also found for methanes adsorbed in a large nano-pore of pore radii R=18.55$\angstrom$ at 105K and 120K. Ball and Evans\cite{ball1988structure} applied the mean-field smoothed density approximation (SDA) density functional theory to study the layering transition of Yukawa fluids under a single wall and cylindrical walls of van der Waals potential. They concluded that the layering transition predicted by density functional theory occur at a temperature below the triple point. A similar argument is made by Peterson et al\cite{peterson1986fluid} in a study of LJ fluids adsorption in a cylindrical pores of smeared-out LJ 9-3 potentials by using mean-field density functional theory. They found  discontinuous first-order layering transition happened if it is below bulk triple point, and the sharp layering transition degrades into a smoothed transition at triple point. Discontinuous layering transition at temperatures above triple points, however, have been reported from well-designed experimental studies of methanes adsorption\cite{inaba1986multilayers} and ethylene adsorption on graphite\cite{drir1986multilayer}. In our study, sharp first-order layering transitions are found in Fig.~\ref{fig:isotherm_R5}. We attribute the predicted layering transition above tirple point in this study to the non-mean-field dispersion free energy functional and the application of cylindrical Steel potential.

The bulk critical point of methane is 190.6K. Fig.~\ref{fig:isotherm_R3.5} and Fig.~\ref{fig:isotherm_R5} show that the critical points for methane under a cylindrical nano-pore of R=11.13$\angstrom$ and R=18.55$\angstrom$ are much lower than methane's bulk critical point. We here made no attempts to locate the exact critical points for methanes adsorbed in graphite cylindrical pores. However, the impact of nano-pores curvature onto critical points is immediate from a comparison to adsorption in slit-pore. Liu et al stuided the critical points of methane adsorption in slit-like pores under Steel potential by iSAFT, and compared the predicted critical points with GCMC simulation\cite{liu2017adsorption}. For methane in slit-like pore of width 2nm, the critical point predicted by iSAFT is 158K. In Fig.~\ref{fig:isotherm_R3.5} (a), methane adsorbed in a cylindrical pore of diameter 2.2nm at 150K is in a supercritical state. The lowered critical point for methane adsorption in cylindrical pore is due to stronger eccentric confinement.

\subsection{Modeling Bottlebrush Polymers}
Polymer brushes is a class of well-designed polymers that are grafted by one end to surfaces of various shapes. Bottlebrush polymer refers to 1D polymer brushes grafted to a cylindrical backbone.\cite{zhang2005cylindrical} The well-studied synthesis for bottlebrush polymer and its enriched architecture lead to many applications. For example, Djalali et al used nanogold as a backbone for bottlebrush polymers and the hybrid metalic and cylindrical polymer brushes can be used as nanowires\cite{djalali2002amphipolar}. Zhang et al grafted amphaphilic polymer brushes to a backbone of carboxylate groups. Because of the affinity of carboxylate groups to metal ions, this bottlebrush polymer can be used as a cylindrical molecular nanoreactors\cite{zhang2004superparamagnetic,zhang2004template}. 

Modeling the bottlebrush polymers is a challenging task because of the length-scale of the system. Self-consistent field theory (SCFT) is a popular theoretical tool for modeling bottlebrush polymers. Unlike density functional theory, none of the variations of SCFT\cite{szleifer1996tethered, scheutjens1979statistical} can capture local packing effect and layering effect, which are important for understanding solvent distributions and solvation forces\cite{binder2017polymer}. DFT have been used to model the structure of plannar polymer brushes\cite{gong2011solvent,gong2012response,gong2013modeling} and spherical polymer brushes\cite{lo2010spherical}. Non-negligible numerical efforts are required for modeling bottlebrush polymers by DFT because of cylindrical geometry. Here we use iSAFT to model bottlebrush polymers in implicit solvents. Segment interact with each other by hard-sphere repulsion and tangential bonding. The details of molecular dynamics simulation that we compare theoretical results with are given in Appendix \ref{simulations}. 


The iSAFT free energy functional of bottlebrush polymers in this study have ideal contribution, hard-sphere contribution and chain contribution which are given in section II. The contribution of solvents and dispersion will be subjects of future work. To tether the polymers onto a cylindrical backbone, we need to add an external potential. For the end segment that is grafted to a backbone, it is under an external potential:

\begin{gather}
	V_1^{ext}(r) = \left\{
	\begin{array}{lr}
	v & \text{if} \quad r = (R+\sigma_\alpha/2)\\
	\infty &  otherwise,
	\end{array}
	\right.
\end{gather}
and for the other segments:
\begin{gather}
	V_\alpha^{ext}(r) = \left\{
	\begin{array}{lr}
	\infty & \text{for} \quad r \leq (R+\sigma_\alpha/2)\\
	0&  \text{otherwise},
	\end{array}
	\right.
\end{gather}
where $R$ is the backbone radius of bottlebrush polymers, and $\sigma_{\alpha}$ is the segment diameter of polymers. The segment diameter is not temperature dependent since the system is athermal.

The density profile for end segment can be calculated from (\ref{eq:EL2}):
\begin{gather}
	\rho_{1}(0) = \exp(\beta\mu_M)\exp[D_1(0)-\beta v ]I_{1,1}(0)I_{2,1}(0),\nonumber
\end{gather}
and for the other segments:
\begin{gather}
\rho_{\alpha}(r_\alpha) = \exp(\beta\mu_M)\exp[D_1(0)-\beta v ]\exp[D_\alpha(r_\alpha)]I_{1,\alpha}(r_\alpha)I_{2,\alpha}(r_\alpha),\nonumber\nonumber\\
\text{for} \quad r_\alpha > R+\frac{\sigma_\alpha}{2}.\nonumber
\end{gather}

Chain connectivity for the all the segments of bottlebrush polymers are modeled by $I_1$ and $I_2$ integrals in section II except that:
\begin{gather}
	I_{1,1}(r_1) = 1,\nonumber\\
	I_{1,2}(r_2) = \Delta^{1,2}(0, r_2) \quad \text{for} \quad r_2 > R+\frac{\sigma_2}{2},\nonumber\\
	I_{1,\alpha}(r_{\alpha}) = \int I_{1,\alpha-1}(r_{\alpha-1}) \exp[{D_{
	\alpha-1}(r_{\alpha-1})}]\Delta^{{\alpha-1}, \alpha}(r_{\alpha-1}, r_\alpha)dr_{\alpha-1}, \nonumber\\
\text{for} \quad r_\alpha > R+\frac{\sigma_\alpha}{2},\nonumber
\end{gather}
and 

\begin{gather}
I_{2,m}(r_\alpha) = 1,\nonumber\\
I_{2,\alpha}(r_{\alpha}) = \int I_{2,\alpha+1}(r_{\alpha+1}) \exp[{D_{
	\alpha+1}(r_{\alpha+1})}]\Delta^{{\alpha}, \alpha+1}(r_{\alpha}, r_\alpha+1)dr_{\alpha+1}, \nonumber\\
\text{for} \quad r_\alpha > R+\frac{\sigma_\alpha}{2},\nonumber\\
I_{2,1}(0) = \int I_{2,2}(r_{2}) \exp[{D_{
		2}(r_{2})}]\Delta^{{1}, 2}(0, r_2)dr_{2}, \nonumber\\
	\text{for} \quad r_2 \geq R+\frac{\sigma_2}{2}.\nonumber
\end{gather}
The effective chemical potential for the first segment is calculated from the grafting density $\rho_g$ of bottlebrush polymers:

\begin{gather}
	\rho_g = \int dr_1\rho_1(r_1)r_1 = \exp[{\beta \mu_M}] \exp[{D_1(0) - \beta v}] I_{1,1}(0)I_{2,1}(0)
	\label{eq:chemical_potential}
\end{gather}
then density profiles for all segments are:

\begin{gather}
	\rho_{\alpha}(r_\alpha) = \frac{\rho_g}{I_{1,1}(0)I_{2,1}(0)}\exp[D_\alpha(r_{\alpha})]I_{1,\alpha}(r_{\alpha})I_{2,\alpha}(r_\alpha)\nonumber\\
	\text{for} \quad r_\alpha > R+\frac{\sigma_\alpha}{2}.
\end{gather}

\begin{figure}[!t]
	\centering
	\includegraphics[scale=0.6]{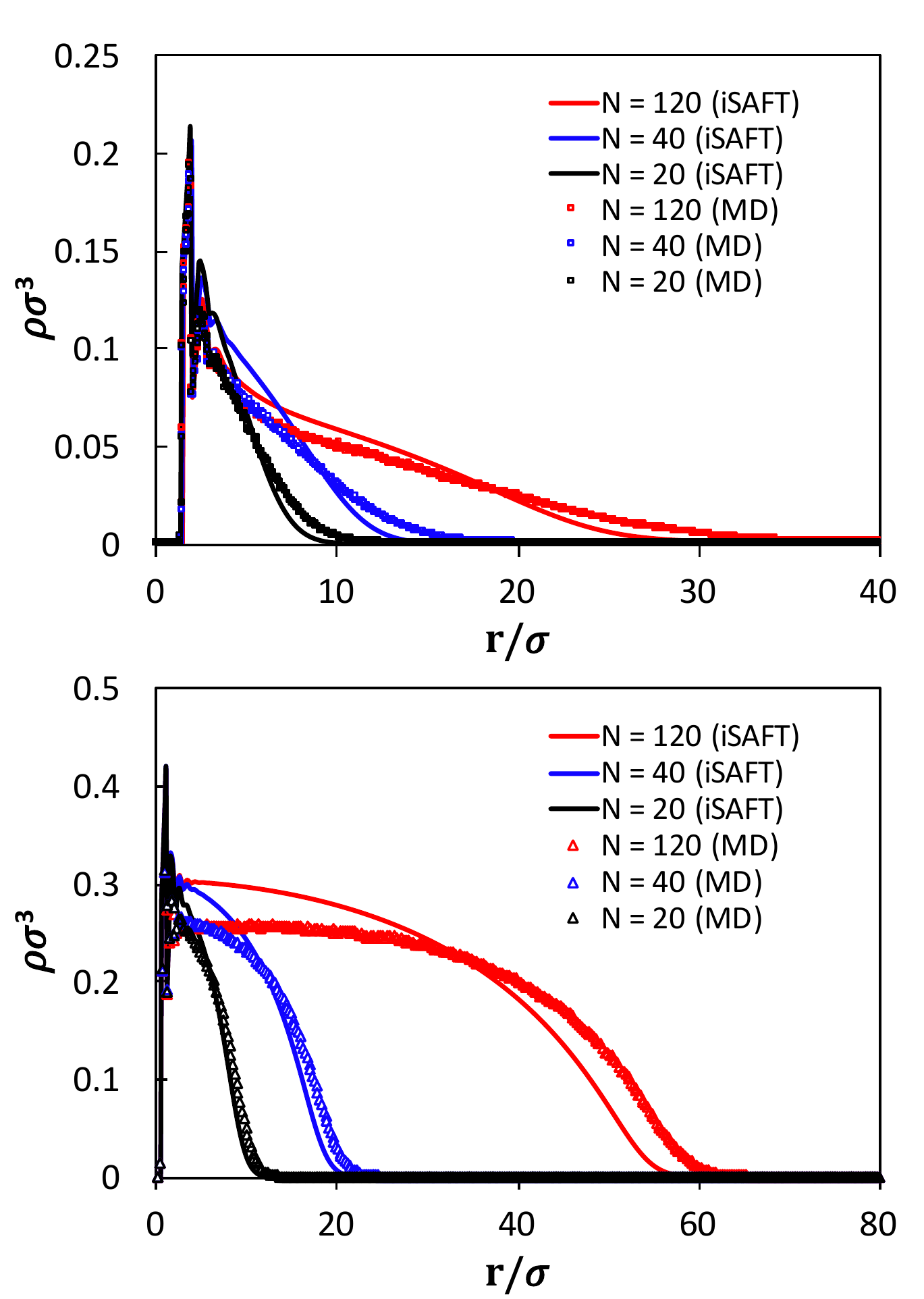}
	\caption{Segment density profiles of hard-sphere chains of varying lengths tethered to surface of different geometries. (a) polymerbrush tethered to cylindrical surface. (b) polymer brush tethered to flat surface. Solid lines are iSAFT predictions.  ($\Box$) are MD simulation results for cylindrical surface. ($\vartriangle$) are MD simulation results for flat surface. Chain length N varies from 20 to 120.}
	\label{fig:chain}
\end{figure}

Fig.~\ref{fig:chain} shows the distribution of polymer brushes grafted to a planar surface (a) and to a cylindrical surface for bottlebrush polymers (b). All the radii of the bottlebrush polymers $R^*$ is set to $1$. A small radii means a large curvature which can effectively differentiate the bottlebrush polymers in Fig.~\ref{fig:chain}(a) from planar polymer brushes in Fig.~\ref{fig:chain}(b). The grafting densities of (a) and (b) are fixed at $\rho_g^* = 0.1$. The chain lengths of polymer brushes range from 20 to 120. A power law shape is observed for bottlebrush polymer profile and a parabolic shape is observed for polymer grafted to a flat surface. This also agrees with the study by Wwijmans et al\cite{wijmans1993polymer} using analytical self-consistent field model (SCF). The curvature effects from the surface shape leads to different levels of repulsive force exerted onto monomers. The spans of polymer brush density distributions as seen in Fig.~\ref{fig:chain} (a) and (b) imply different scaling laws for radius of gyration, which is confirmed by other studies\cite{milner1988theory, wijmans1993polymer}.
 
\begin{figure}[!t]
	\centering
	\includegraphics[scale=0.5]{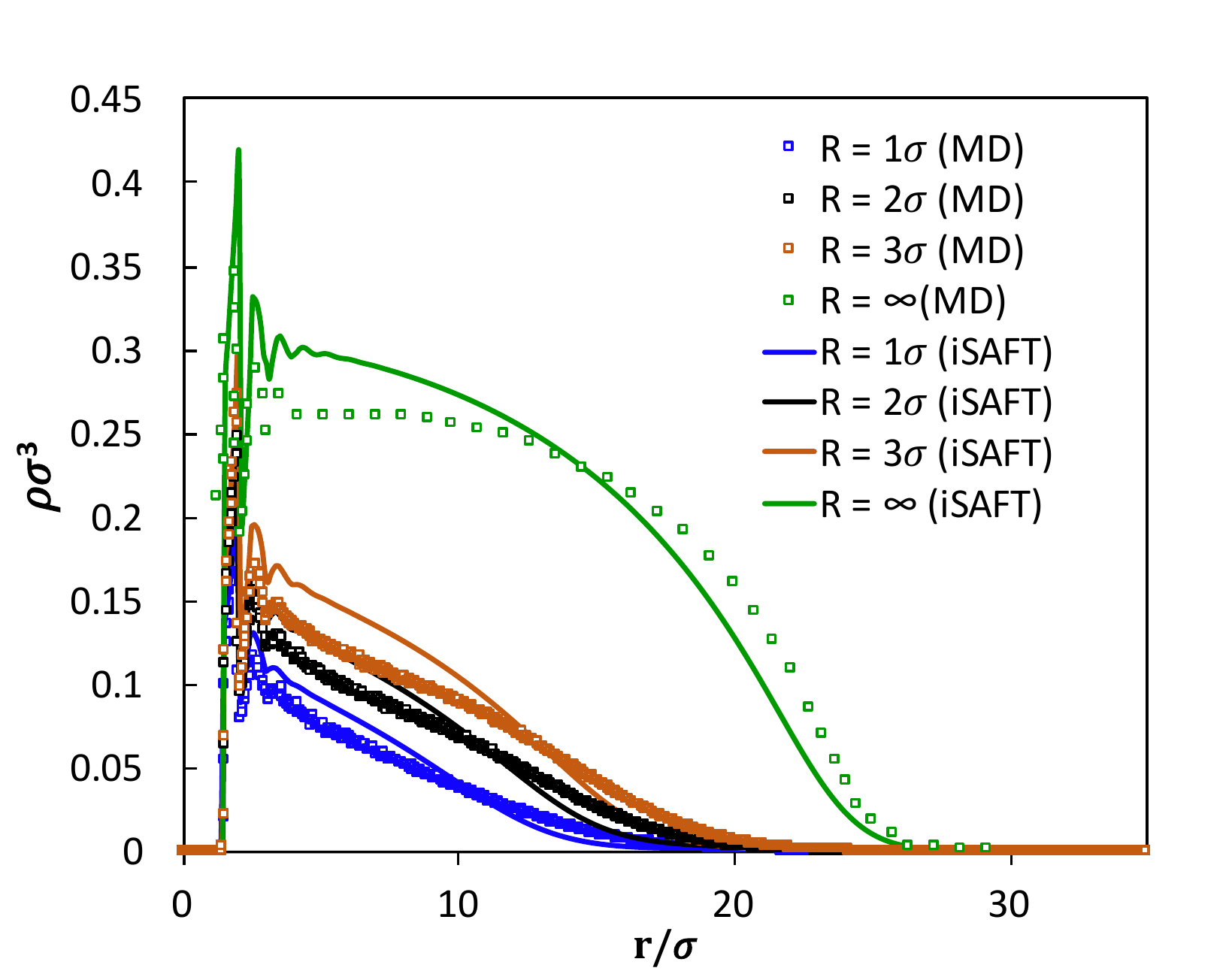}
	\caption{Segment density profiles of bottlebrush polymers and planar polymerbrushes. $R$ stands for the backbone radius of bottlebrush polymers. Bottlebrush polymers are in blue, black and brown. Planar polymerbrushes are in green. Solid lines represent iSAFT results. ($\Box$) represent molecular dynamics simulation results. Details of MD simulation are provided in Appendix \ref{simulations}. MD simulation of planar polymerbrushes are from the study by Grest et al\cite{grest1993structure}.}
	\label{fig:curvature}
\end{figure}

Fig.~\ref{fig:curvature} shows how the microstructures of polymer brushes vary for different curvatures at fixed grafting density and fixed chain length. As the radius of bottlebrush backbone increases, a continual transition from a power law to parabolic distribution is observed. Flat surface can treated as a limiting case of infinitely large curvature. The density of the adsorption polymers close to backbones also increases because larger radii result in stronger entropic repulsive forces. It can be seen that the convergence process towards a zero curvature flat wall is very slow. This is also found by Roth et al in a study of depletion force exerted by curved surfaces\cite{roth1999depletion}.

\section{Conclusion}
iSAFT has been successful in modeling the behaviors of inhomogeneous fluids. Numerical challenge however existed for the application of iSAFT in fluid in cylindrical geometry. We have presented an efficient numerical solution method in resolving this problem by using fast Hankel transform on a non-equidistant grid. Improvement in numerical efficiency and time scaling are achieved. The two applications of the methods are discussed. Interesting physical insights are obtained from the applications that theory predicts layering transitions above the triple points for methane adsorptions, and theory captures power-law to parabolic transitions of polymer brush microstructures.

\appendix
\section{Molecular Simulation Details}
\label{simulations}

\subsection{Grand Canonical Monte Carlo Simulation}
In order to test the performance of the weighted approximation approach to model fluids under cylindrical geometry, we perform grand canonical Monte Carlo (GCMC) simulations of Lennard--Jones spheres under a cylindrical Steele potential~\cite{siderius2011extension} to model methane adsorption in cylindrical nano-pores.

During each Monte Carlo run, the simulation attempts to either displace a particle to change the system configuration or exchange a particle with constant chemical potential bath under equal probability. Particle displacement is accepted with probability: 

\begin{equation*}
acc(\mathbf{r}_N\rightarrow\mathbf{r}_N') = \min\left[1, \exp\left(\varDelta U\right/k_BT)\right],
\end{equation*}
where $\varDelta U$, is the change of configuration energy if a particle displacement is accepted.

The exchange of a particle is accepted at a criterion related to bulk chemical potential. A particle insertion is accepted with probability:

\begin{equation*}
acc(N \rightarrow N+1) = \min\left[ 1, \frac{zV\exp\left(\varDelta U/k_BT\right)}{N+1}  \right].
\end{equation*}
A particle removal is accepted with probability:
\begin{equation*}
acc(N \rightarrow N-1) = \min\left[ 1, \frac{N\exp\left(\varDelta U/k_BT\right)}{zV}  \right].
\end{equation*}

Here, V is the system volume, $\varDelta U$ is the change of configuration energy after particle exchange, z is the absolute fugacity that relates density functional theory to GCMC. The activity can be found from the relation:

\begin{equation*}
z = \rho_B\exp\left[-c^{(1)}\left(\rho_B\right)\right],
\end{equation*}
where $c^{(1)}\left(\rho_B\right)$ is the one-body direct correlation function of the bulk fluid at density $\rho_B$. This can be obtained by taking functional derivative of (\ref{eq:FMT}) and (\ref{eq:WDA}) and taking it to bulk limit. Alternative, it can be calculated from PC-SAFT equation of state at the same bulk density and parameters.

The parameters for methane in GCMC is from the TraPPE Force Field parameters~\cite{martin1998transferable}. The system is equilibrated after $4\times10^6$ Monte Carlo moves which include configuration change and particle insertion. The parameter of maximum particle displacement is adjusted that the overall acceptance rate for configuration change is $50\%$. After the equilibration, another $1\times10^6$ Monte Carlo moves are generated to sample the fluid distribution every 10 Monte Carlo moves.

\subsection{Molecular Dynamics Simulation}
Testing of the algorithm for single polymer brushes were done against Molecular dynamics data. The simulations were done using LAMMPS molecular dynamics code\cite{plimpton:jcp95}\cite{LAMMPS}. The polymer chains were made of bead-spring freely jointed monomers. The beads are bonded using harmonic potential. The non-bonded interactions were defined by truncated and shifted 12-6 Lennard-Jones potential. The polymer chains were then grafted onto the required surface (planar or cylindrical) by fixing the first bead at randomly onto the surface at a given density, and the rest of the beads grown outwards. The surface was emulated using a wall potential of cylindrical and planar surface. The beads interacted with the surface with the same truncated and shifted 12-6 Lennard-Jones potential. For all monomers, $\epsilon = 1$ and $\sigma = 1$. The cutoff distance was $r_c = 2^{1/6} \sigma$, to emulate hardsphere interactions. The spring constant for the bond was $K=1662 \epsilon / \sigma^2$, with a bond length of $r_0=\sigma$. 

The simulation box was periodic in all dimensions. To ensure that the polymer does not interact with it's periodic image, the radial direction needs to have at least  $20 \sigma$ of space before the cell boundary. This is the $xy$ dimensions for the cylindrical polymer, and the $z$ dimension for planar polymer. The other dimension(s) were set as $40 \sigma$. The time step was chosen as  $\Delta t = 0.0012 \tau$, where $\tau = \sigma \sqrt{m/\epsilon}$. The temperature $T^* = k_B T / \epsilon$  was controlled using Nos\'e-Hoover thermostat with a damping factor of $3\tau$. All simulations were minimized using conjugate gradient method and initialized with at a temperature of $T^* = 6.5$. The system was then cooled to $T^* = 1.0$ in $5 \times 10^5$ time steps, and further relaxed at the same temperature for $5 \times 10^5$ time steps. The thermostat was then turned off and a production run was carried out for $1 \times 10^6$ steps, saving frames every 1000 time steps.

The truncated and shifted 12-6 Lennard-Jones potential is given by
\begin{equation}
u^{LJ}(r) = 4 \epsilon \Bigg [ \Big ( \frac{\sigma}{r} \Big )^{12} - \Big ( \frac{\sigma}{r} \Big )^6 \Bigg ]
\end{equation}
where $\epsilon$ is the well depth, $\sigma$ is the particle diameter and $r$ is the distance.  
The monomer-surface interactions is given as
\begin{equation}
u_{rs}(r_{iz}) = u^{LJ}(r_{iz}-0.5) - u^{LJ}(2^{1/6} - 0.5)
\end{equation}
where $r_{iz}$ is the distance of the center of the bead from the surface. For the cylindrical surface, the distance is from the axis of the rod, i.e. $z$-axis. Fro the planar surface, this is the $z=0$ plane.

The monomer-monomer interactions are given as
\begin{equation}
u_{ss}(r_{ij}) = u^{LJ}(r_{ij}) - u^{LJ}(r_c)
\end{equation}
where $r_{ij}$ is the distance between the centers of the two beads.

The bonded monomer interactions is represented by a harmonic potential. 
\begin{equation}
u^{bond}(r_{ij}) = K(r_{ij} - r_0)^2
\end{equation}
where $K$ is the spring constant and $r_0$ is the equilibrium bond length.  

\section{Free Energy Functional Derivatives for Hard Sphere Contribution in Cylindrical Geometry}
\label{derivatives}
Rosenfeld's free energy functional agrees with the Percus$-$Yevick compressibility equation of state in the bulk limit. The free energy functional density $\Phi$ includes four scalar weighted densities and two vector densities. The four scalar weighted densities can be reduced to two types of convolutions. The two vectorial densities can be reduced to one type of convolution. Those weighted densities enable us to represent the vectorial densities by two scalar weighted densities, thus conserving the convolution properties of FMT in cylindrical geometry.

The FMT free energy functional for hard-sphere fluid is:
\begin{equation*}
A^{hs}[\rho] = k_BT\int \Phi(\mathbf{r}) d\mathbf{r},
\end{equation*}
where  $\Phi$ is:

\begin{align}
\Phi(\mathbf{r})&=-n_0(\mathbf{r})\ln(1-n_3(\mathbf{r}))\nonumber\\
&+\frac{n_1(\mathbf{r})n_2(\mathbf{r})-\mathbf{n}_{v1}(\mathbf{r})\cdot\mathbf{n}_{v2}(\mathbf{r})}{1-n_3(\mathbf{r})}\nonumber\\ 
&+\frac{n_2(\mathbf{r})^3-3n_2(\mathbf{r})\mathbf{n}_{v2}(\mathbf{r})\cdot\mathbf{n}_{v2}(\mathbf{r})}{24\pi(1-n_3(\mathbf{r}))^2}.
\label{eq:FMT}
\end{align}
Since we are dealing with mixtures, each of the weighted densities are averaged over all $m$ segments:
\begin{align}
n_i(\mathbf{r}) &= \sum_{\alpha=1}^{m}n_{i,\alpha}(\mathbf{r})\nonumber\\ &= \sum_{\alpha=1}^{m} \rho_{\alpha}(\mathbf{r}') *\omega_{\alpha}^{(i)}(\mathbf{r} - \mathbf{r}') ,    \quad \text{for} \quad i=0,1,2,3,v1,v2,
\label{eq:weighted_1}
\end{align}
where the weights $\omega_{\alpha}^{(i)}$ are given as:

\begin{align}
	\omega_{\alpha}^{(0)}(\mathbf{r}) = \frac{\delta(R_\alpha - |\mathbf{r}|)}{4\pi R_\alpha^2}\\
	\omega_{\alpha}^{(1)}(\mathbf{r}) = \frac{\delta(R_\alpha - |\mathbf{r}|)}{4\pi R_\alpha}\\
	\omega_{\alpha}^{(2)}(\mathbf{r}) = \delta(R_\alpha - |\mathbf{r}|)\\
	\omega_{\alpha}^{(3)}(\mathbf{r}) = \Theta(R_\alpha - |\mathbf{r}|)\\
	\omega_{\alpha}^{(v1)}(\mathbf{r}) = \frac{\mathbf{r}}{|\mathbf{r}|}\delta(R_\alpha - |\mathbf{r}|)\\
	\omega_{\alpha}^{(v2)}(\mathbf{r}) = \frac{\mathbf{r}}{|\mathbf{r}|}\frac{\delta(R_\alpha - |\mathbf{r}|)}{4\pi R_\alpha}
\end{align}
where $\ast$ denotes spatial convolution. 

The functional derivatives of FMT is:
\begin{align}
	\frac{\delta A^{hs}[\rho]/k_BT}{\delta\rho_{\alpha}(\mathbf{r})} =  \sum_{i=0,1,2,3,v1,v2} \frac{\partial\Phi}{\partial n_{i}(\mathbf{r}')} *\omega_{\alpha}^{(i)}(\mathbf{r} - \mathbf{r}')\nonumber\\  \quad \text{for} \quad i=0,1,2,3,v1,v2.
	\label{eq:hs_derivative}
\end{align}
%
%

Under the assumption that there is a uniform cylindrical external potential i.e., $V_\alpha^{ext}(z,\theta,r) = V_\alpha^{ext}(r)$, and $r = \sqrt{x^2+y^2}$, the three-dimensional microscopic density distribution $\rho_{\alpha}(\mathbf{r})$ reduces to a one-dimensional density distribution $\rho_{\alpha}(r)$, so do other scalar weighted densities $n_{0,\alpha}(r)$, $n_{1,\alpha}(r)$, and $n_{2,\alpha}(r)$. The weighted vectorial densities $n_{v1,\alpha}(\mathbf{r})$ and $n_{v2,\alpha}(\mathbf{r})$ reduce to one-dimensional vectors $n_{v1,\alpha}(r) \hat{e}_r$ and $n_{v2,\alpha}(r) \hat{e}_r$ in cylindrical geometry. $\hat{e}_r$ is the unit vector in radial direction.

The scalar parts of the vectorial weighted densities $n_{v1,\alpha}(r)$ and $n_{v2,\alpha}(r)$ are more involved. We consider the procedure in the reference\cite{gonzalez2006density}:

\begin{equation}
	n_{v1,\alpha}(r) = \frac{1}{2r}\left[\left(r^2+R_{\alpha}^2\right)n_{0,\alpha}(r) - \tau_{\alpha}(r) - \frac{n_{3,\alpha}(r)}{4\pi R_\alpha}
	\right],
\end{equation}
where $\tau_{\alpha}(r)$ has to be calculated by another convolution:

\begin{equation*}
\tau_{\alpha}(r) = \frac{1}{4\pi R_{\alpha}^2}\int_{}^{}{r'}^2\rho_{\alpha}(r')\delta(R_{\alpha}-|\mathbf{r}-\mathbf{r}'|)d\mathbf{r}'.
\end{equation*}

\section{Fourier Transform of Weighting Functions in Cylindrical Geomtry}
\label{transform}
The Fourier transforms of weighting functions have closed-form solutions. Here we provide their derivations. 


\begin{align}
\tilde{\omega}_{\alpha}^{(0)}(\mathbf{k})& =  \mathcal{F}\{\omega_{\alpha}^{(0)}(\mathbf{r})\} \nonumber\\&=  \int_{-\infty}^{\infty}\frac{1}{4\pi R_{\alpha}^2}\delta(R_{\alpha}-|\mathbf{r}|){\rm e}^{-i\mathbf{k}\cdot \mathbf{r}} d\mathbf{r} \nonumber\\&= \frac{1}{4\pi R_{\alpha}^2}
\int_{0}^{2\pi}d\phi\int_{0}^{\pi}\int_{0}^{\infty}\delta(R_{\alpha}-r){\rm e}^{-ikr\cos\theta}r^2\sin\theta dr d\theta\nonumber\\&= \frac{1}{2}
\int_{-1}^{1}	{\rm e}^{-ikR_\alpha\cos\theta}d\cos\theta \nonumber\\&=	 \frac{-{\rm e}^{-ikR_\alpha\cos\theta}}{2ikR_\alpha}\bigg|^{1}_{-1} \nonumber\\&=
\frac{\sin(kR_\alpha)}{kR_\alpha}.
\end{align}

We arrive at the third line by first reformulating the transform in a spherical coordinate then set the direction and origin of $\mathbf{k}$ vector the same as $z$ axis of the spherical coordinate. $|\mathbf{r}| = r$ and $|\mathbf{k}| = k$ due to the cylindrical symmetry. 

\begin{align}
\tilde{\omega}_{\alpha}^{(3)}(\mathbf{k})& =  \mathcal{F}\{\omega_{\alpha}^{(3)}(\mathbf{r})\} \nonumber\\&=  \int_{-\infty}^{\infty}\Theta(R_{\alpha}-|\mathbf{r}|){\rm e}^{-i\mathbf{k}\cdot \mathbf{r}} d\mathbf{r} \nonumber\\&= 4\pi
\int_{0}^{\infty}\Theta(R_{\alpha}-r)\frac{\sin(kr)}{k} rdr\nonumber\\&= 
\frac{4\pi}{k^3}\left[\sin(kR_\alpha) - kR_\alpha\cos(kR_\alpha)\right].
\end{align}

%

The calculation of weighted density and functional derivative of dispersion term requires a transform of body average weighting function. 

\begin{align}
\tilde{\omega}_\alpha^{avg}(\mathbf{k})& =  \mathcal{F}\{\omega_{\alpha}^{avg}(\mathbf{r})\} \nonumber\\&=  \int_{-\infty}^{\infty}\frac{3}{4\pi d^3_\alpha}\Theta(d_{\alpha}-|\mathbf{r}|){\rm e}^{-i\mathbf{k}\cdot \mathbf{r}} d\mathbf{r} \nonumber\\&= \frac{3}{d^3_\alpha}
\int_{0}^{\infty}\Theta(d_{\alpha}-r)\frac{\sin(kr)}{k} rdr\nonumber\\&= 
\frac{3}{d^3_\alpha k^3}\left[\sin(kd_{\alpha}) - kR_\alpha\cos(kd_{\alpha})\right].
\end{align}

One should be aware of the limiting case of those transforms when frequency $k\rightarrow0$. The limiting values are obtained using L'Hospital's Rule:

\begin{flalign}
\lim_{k\rightarrow0}\tilde{\omega}_{\alpha}^{(0)}(\mathbf{k}) &= \lim_{k\rightarrow0}\cos(kR_\alpha) = 1.&
\end{flalign}

\begin{flalign}
\lim_{k\rightarrow0}\tilde{\omega}_{\alpha}^{(3)}(\mathbf{k}) &= \lim_{k\rightarrow0}\frac{4\pi}{3k^2}\left[R_\alpha\cos(kR_\alpha) - R_\alpha\cos(kR_\alpha) + kR^2_\alpha\sin(kR_{\alpha}) \right] \nonumber\\&= \lim_{k\rightarrow0}\frac{4\pi}{3k}\left[R^2_\alpha\sin(kR_\alpha)\right] \nonumber\\&=\frac{4\pi R^3_{\alpha}}{3}.
\end{flalign}


\begin{flalign}
\lim_{k\rightarrow0}\tilde{\omega}_{\alpha}^{avg}(\mathbf{k}) & = 1.&
\end{flalign}

\begin{acknowledgments}
The authors thank the Robert A. Welch Foundation (Grant No. C-1241) for financial support.
\end{acknowledgments}

\nocite{*}
\bibliography{algorithm}

\end{document}